\documentclass[aoas]{imsart}

\RequirePackage{amsthm,amsmath,amsfonts,amssymb}
\RequirePackage{dsfont}
\RequirePackage[authoryear]{natbib}
\RequirePackage{graphicx,ulem}
\RequirePackage{rotating}
\RequirePackage{subcaption}
\RequirePackage{enumerate}
\RequirePackage{bm}
\RequirePackage[colorlinks,citecolor=blue,urlcolor=blue]{hyperref}

\startlocaldefs
\theoremstyle{plain}

\endlocaldefs

\begin{document}

\begin{frontmatter}
\title{Modeling and Simulating Dependence in Networks Using Topological Data Analysis}
\runtitle{Modeling and Simulating Dependence in Networks}

\begin{aug}
\author[A]{\fnms{Anass} \snm{El Yaagoubi Bourakna}\ead[label=e1,mark]{anass.bourakna@kaust.edu.sa}},
\author[B]{\fnms{Moo} \snm{K. Chung}\ead[label=e2,mark]{mkchung@wisc.edu}}
\and
\author[A]{\fnms{Hernando} \snm{Ombao}\ead[label=e3,mark]{hernando.ombao@kaust.edu.sa}}
\address[A]{King Abdullah University of Science and Technology (KAUST), Statistics Program, CEMSE Division \printead{e1}, \printead{e3}}
\address[B]{Department of Biostatistics \& Medical Informatics, University of Wisconsin–Madison \printead{e2}}

\end{aug}

\begin{abstract}
    Topological data analysis (TDA) approaches are becoming increasingly popular for studying the dependence patterns in multivariate time series data. In particular, various dependence patterns in brain networks may be linked to specific tasks and cognitive processes, which can be altered by various neurological and cognitive impairments such as Alzheimer's and Parkinson's diseases, as well as attention deficit hyperactivity disorder (ADHD). Because there is no ground-truth with known dependence patterns in real brain signals, testing new TDA methods on  multivariate time series is still a challenge. Simulations are crucial for evaluating the performance of proposed TDA methods and testing procedures as well as for creating computation-based confidence intervals. To our knowledge, there are no methods that simulate multivariate time series data with specific and manually imposed connectivity patterns. In this paper we present a novel approach to simulate multivariate time series with specific number of cycles/holes in its dependence network. Furthermore, we also provide a procedure for generating higher dimensional topological features.
\end{abstract}


\end{frontmatter}



\section{Introduction}
\label{sec:introduction}

Topological data analysis (TDA) has witnessed many important advances over the last twenty years that aim to unravel and provide insight to the "shape" of the data (\cite{TDA_EDELSBRUNNER}, \cite{BARCODES} and \cite{EDELSBRUNNER_HARER}). The development of TDA tools such as barcodes and persistence diagrams have opened many new perspectives for analyzing various types of data. These tools help the practitioner to understand the topological features present in high dimensional data, which are not directly accessible using other classical techniques.

Many data sets have a temporal structure (e.g., financial time series, climate data, brain signals). In recent years, there has been a shift from employing TDA techniques on clouds of points to applying them on dependence networks of multivariate time series data, particularly for multivariate brain signals such as electroencephalograms (EEG) and local field potentials (LFP), see \cite{TDA_MULTIVARIATE_TS_ANASS}. Rather than using TDA techniques on a cloud of points via a time delay embedding transformation, it is suggested that the multivariate time series be transformed to its dependence network where the nodes correspond to the time series components and the weight on the edges depend on the intensity of the statistical dependency between any given pair of time series in a network. There is currently no systematic method nor statistical model for conducting simulations with complex dependence structure in a network. This is a serious limitation because these TDA methods cannot be evaluated for sensitivity, specificity, predictive ability. In this paper, we will develop an easily implementable method for simulating data with complex dependence in a network. Thus, the main contributions of this work are (a.) rigorous evaluation of proposed methods and (b.) statistical inference using simulations and resampling methods.

It is known in the literature that the topology of the brain network (structural and functional) is organized according to principles that maximize the flow of information and minimize the energy expenditure for maintaining the entire network, such as small world networks (\cite{STRUCTURE_FUNCTION_BRAIN_NETWORKS}, \cite{BRAIN_NETWORKS_ORGANIZATION} and \cite{SMALL_WORLD_NETWORKS}). This topological structure of the brain network can be altered by various conditions such as attention deficit hyperactivity disorder (ADHD), Alzheimer's and Parkinson's diseases. Topological tools have been developed over the years to assess and analyze the topological patterns of different groups' brain networks (healthy vs pathological), as well as quantifying the impact of these disorders on brain organization. There are many approaches for generating multivariate time series data: parametric vector autoregressive moving average (VARMA) models, \cite{DUAL_FREQUENCY_COHERENCE}; and by stochastic representations in terms of some basis, \cite{SLEX}, there are no known methods for generating multivariate time series data with specific number of cycles in its dependence pattern, making proper testing of these novel topological tools and summaries impossible.

Motivated by the challenge above, we propose a novel approach to generate multivariate time series data with specified number of cycles in its dependence structure. As noted, this is significant because this will meet the need for rigorous evaluation of TDA methods for multivariate time series; and also the need for a proper statistical inference procedure in TDA for multivariate time series. The observed brain signals can be characterized as the superposition or mixture of frequency specific latent sources (\cite{MOTTA_OMBAO}, \cite{AR_MIXTURE_MODELS}, \cite{BINARY_TS_GAUSSIAN_PROCESSES}, \cite{BATESIAN_MODELING_MULTIVARIATE_TS}, \cite{BRAIN_SIGNALS_LOW_EMBEDDING}, \cite{AR_MIXTURES_BRAIN_SIGNALS} and \cite{SPECTRAL_DEPENDENCE}). In this paper, we develop a procedure that is based on the idea of mixing latent second-order autoregressive (AR(2)) processes to generate various patterns of connectivity. In Section 2, we recall some basic TDA concepts. In Section 3, we provide a brief overview of AR(2), then explain how to generate various dependence patterns by mixing such processes using carefully selected weights. In Section 4, we generate multivariate time series data with cyclic dependence structure and show that the persistence diagram detects such structures. In Section 5, we explain how to generate multivariate time series data with more general patterns. In Section 6, we study the sensitivity of our approach to generate patterns in the dependency structure as a function of the signal to noise ratio. In Section 7, we use our approach to carry out simulation-based inference based on the notion of total persistence.

\section{Topological Data Analysis: An overview}
\label{sec:TDA_Overview}

The goal of topological data analysis for time series is to provide computational tools that can assess the topological features present in the dependence network of a multivariate time series' dependence network. Through the use of persistent homology theory, TDA provides a framework for analyzing the topological features, such as connected components, holes, cavities, etc. that are present in the network (\cite{EDELSBRUNNER_HARER} and \cite{PH_CHUNG}).

\begin{figure}[ht]
    \centering
    \includegraphics[width=.7\linewidth]{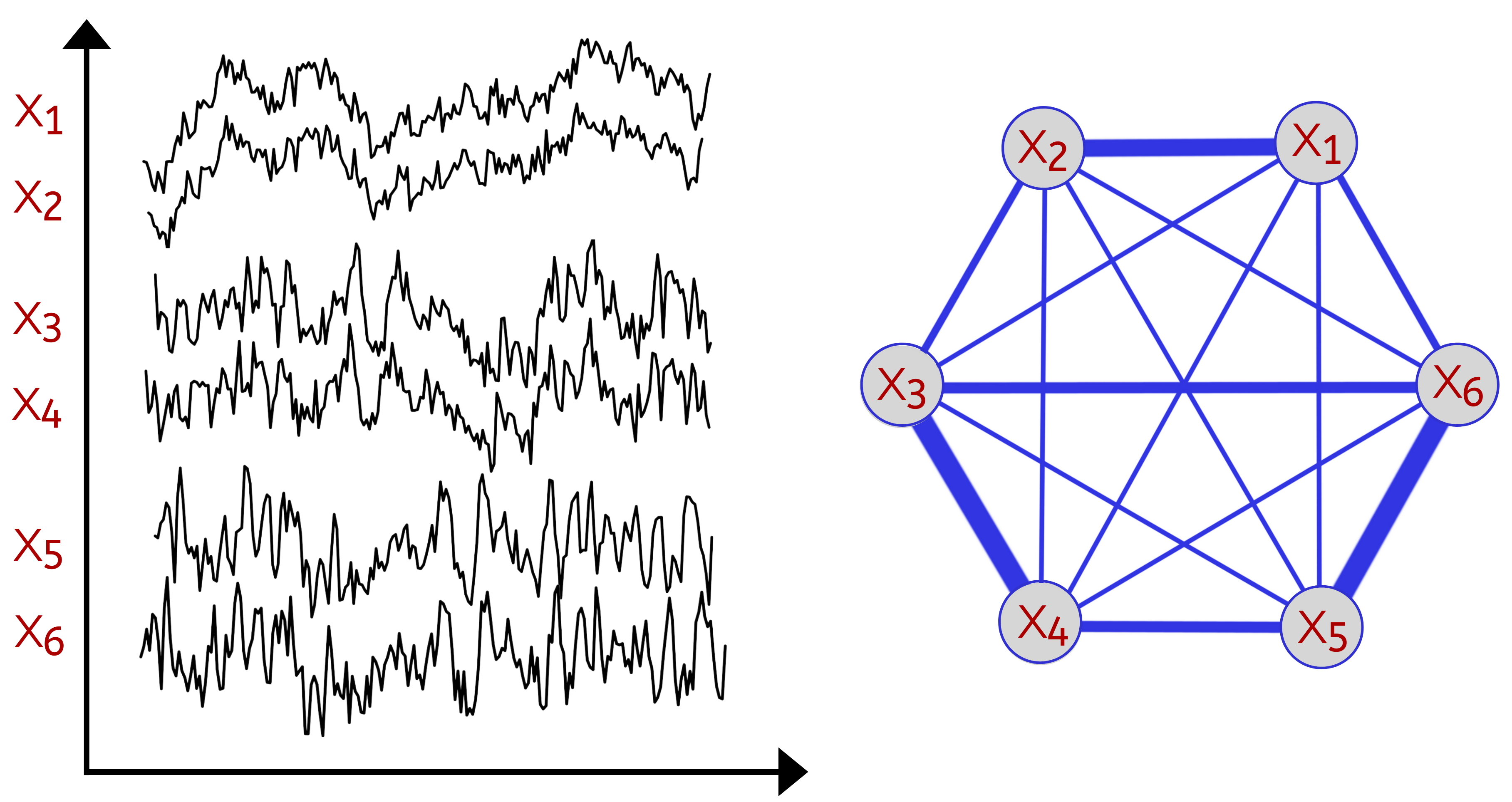}
    \caption{Left: Example of a multivariate time series with P=6 channels. Right: Dependence network of the multivariate time series.}
    \label{fig:TS_Dependence_Network}
\end{figure}

\begin{figure}[ht]
    \centering
    \includegraphics[width=\linewidth,height=1.25in]{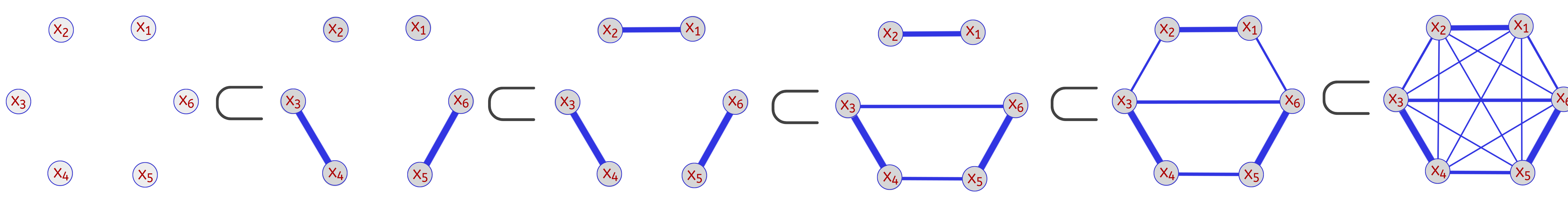}
    \caption{Consider a network with $P$ nodes (brain regions) with corresponding time series $X_1(t), ...  X_P(t)$, we define the distance between two time series components as a decreasing function of dependence. For every threshold $\epsilon$, a network with edges with weights not exceeding $\epsilon$ is constructed leading to an increasing sequence also known as Vietoris-Rips filtration. As the threshold $\epsilon$ grows, there's birth and death of various topological features.}
    \label{fig:VR_Filtration}
\end{figure}

In order to analyze the topological features present in the various dependence networks, we usually consider the homology of the filtration (increasing sequence of thresholded networks) obtained from these dependence networks, also known as Vietoris-Rips filtration. Let $X(t) = [X_1(t), \hdots, X_P(t)]'$ be the observed brain signals at $P$ different locations at time $t \in \{1, ...T \}$. One can define the dependence network to be a weighted graph with weights between nodes $p$ and $q$ to be some dependence measure between the observed time series components $X_p(t)$ and $X_q(t)$, as can be seen in Figure \ref{fig:TS_Dependence_Network}.
Since the Vietoris-Rips filtration relies on the notion of distance, we can then define distance between brain channels at locations $p$ and $q$ to be a decreasing function of the dependence. For example, using coherence or correlation to measure dependence, one can use the transform $g(x) = 1-x$ (where $x$ is correlation or coherence). Therefore, using this measure of distance, the Vietoris-Rips filtration is constructed by connecting nodes that have a distance less or equal to some given threshold $\epsilon$, which results in the following filtration:
\begin{align}
    \mathcal{X}_{\epsilon_1} \subset \mathcal{X}_{\epsilon_2} \subset \cdots \subset \mathcal{X}_{\epsilon_n}, \label{eq:filtration}
\end{align}
where $0 < \epsilon_1 < \epsilon_2 < \cdots < \epsilon_{n-1} < \epsilon_n$ are the distance thresholds, see illustration in Figure \ref{fig:VR_Filtration}. Refer to \cite{HAUSMANN_RIPS_FILTRATION} for a review of Vietoris-Rips filtrations.

\begin{figure}[ht]
    \centering
    \includegraphics[width=.6\linewidth]{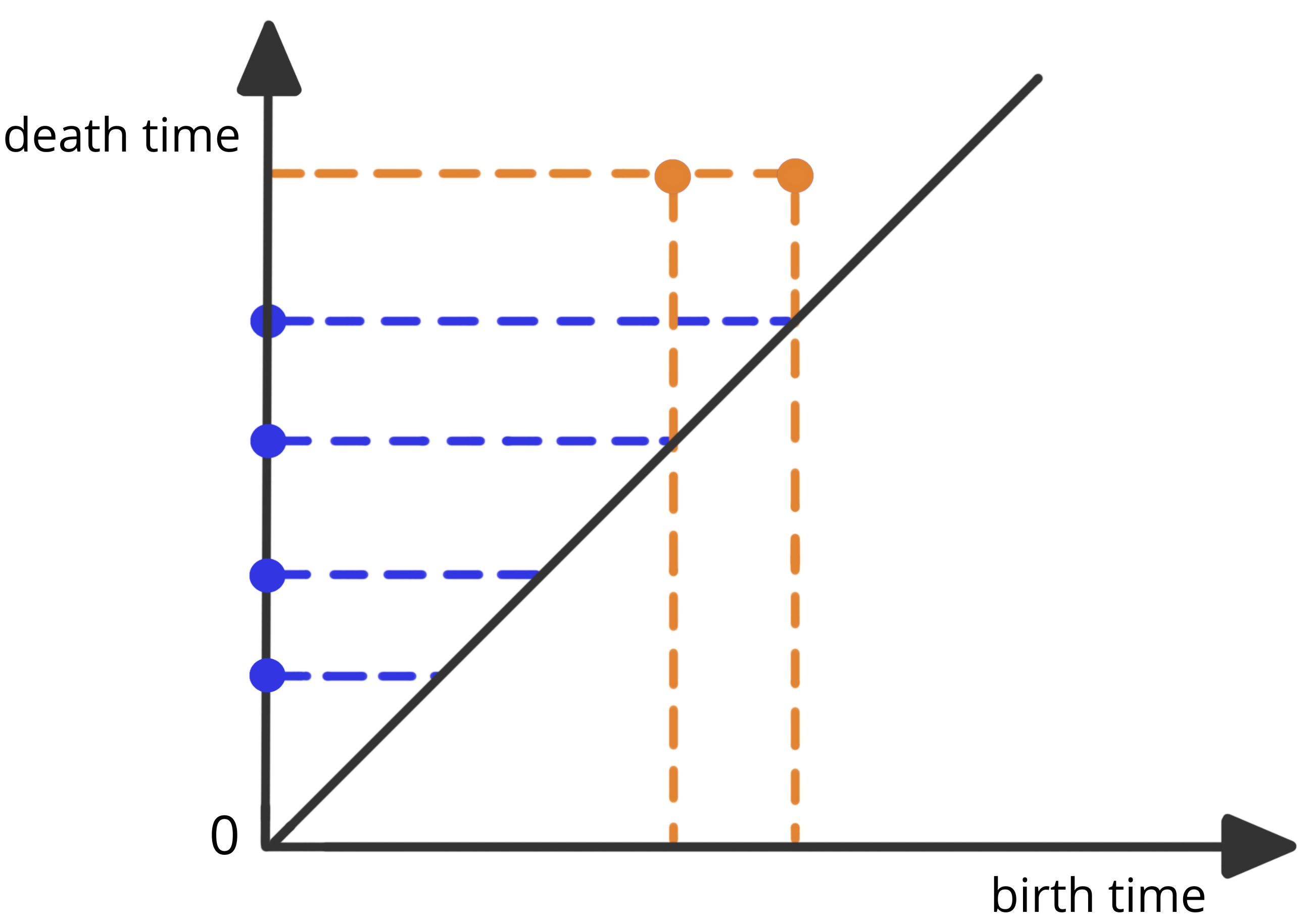}
    \caption{Example of a persistence diagram corresponding to the previous Vietoris-Rips filtration. The various dots correspond to different birth-death pairs ($b_i, d_i$) with birth time in the x-axis and death time in the y-axis.}
    \label{fig:PD_example}
\end{figure}

The objective of this approach is to examine the nature of topological features at different scales as they (connected components, cycles, holes etc.) appear (birth) and then vanish (death) (\cite{TDA_EDELSBRUNNER}, \cite{BARCODES_FIRST}, \cite{EDELSBRUNNER_HARER} and \cite{BARCODES}). The Vietoris-Rips filtration can be a complex object. Therefore, in practice, practitioners consider a topological summary that is known as the persistence diagram (PD), which is a diagram that represents the times of birth and death of the topological features in the VR filtration as seen in Figure \ref{fig:PD_example}. Every birth-death pair is represented by a point in the diagram, e.g., ($b_1$, $d_1$) and  ($b_2$, $d_2$), where $b_1$ is the birth time of the first feature and $d_1$ is the death time of the first feature etc. The points in the PD are colored based on the dimension of the feature they correspond to (e.g., one color for the connected components, another color for the cycles etc.).

\section{Modeling dependence patterns based on mixtures of AR(2) processes}
\label{sec:dependence_patterns}

In neuroscience, the concept of the nervous system as a (structural and functional) network of interconnected neurons is now well established (\cite{NETWORK_FUNCTIONAL_CONNECTIVITY},
\cite{SPORNS_BRAIN_CONNECTIVITY},
\cite{KARL_FRISTON} and
\cite{MULTISCALE_BRAIN_CONNECTIVITY}). Many brain investigations have led to countless discoveries concerning the brain's anatomical and functional organization. The ongoing scientific endeavor in neuroscience to map the complicated networks of the human brain with increasing accuracy thanks to technical advances in brain imaging techniques has resulted in a plethora of new statistical techniques that aim to study and analyze various patterns in these complex networks. Such methods not only help neuroscientists understand the segregation of brain functions but also the integration of information processing. As a result, the validity of such novel techniques must be evaluated in terms of various metrics such as false positive/false negative rates, type I and type II errors etc. It is impossible to evaluate/assess such approaches without a proper method for generating multivariate time series data with ground truth patterns in its dependence network.

Given the intrinsic complexity of brain signals which are sometimes thought of as the superposition of different frequency specific underlying components. It might be difficult to discover and analyze the interrelationships between distinct time series components. As a result, this paper will adopt a frequency-specific strategy to generate meaningful simulations. Thus, we will develop a method where the multivariate time series data with dependency connections that are allowed to vary across frequency bands. For this reason, we will consider coherence as our frequency-specific dependence measure, since it can capture lead-lag dependencies. The typical approach for analyzing brain data requires estimating the spectrum first, then constructing the connectivity network using a spectral dependence measure, usually, coherence or partial coherence
(\cite{COHERENCE_BRAIN_CONNECTIVITY} and \cite{PDC_BRAIN_CONNECTIVITY}). In the following subsections we will recall some important properties of AR(2) processes, then we will explain the key idea behind using mixtures of AR(2) processes to generate various dependence patterns.

\subsection{Autoregressive processes of order 2 (AR(2))}

Electrophysiological signals are viewed as mixtures of autoregressive processes. As a result of the flexibility of AR(2) processes, i.e., their ability to represent oscillations at precise frequency bands, we choose to use these as building blocks to simulate brain signals with various dependency patterns that are frequency specific. Following the idea developed in \cite{EEG_LEAD_LAG_REGRESSION}, \cite{SSM_LFP}, \cite{SPECTRAL_DEPENDENCE} and \cite{AR_MIXTURES_BRAIN_SIGNALS}, that electroencephalograms (EEGs) could be viewed as mixtures of latent frequency specific sources, i.e., mixtures of frequency-specific neural oscillations.

\begin{figure}[ht]
    \centering
    \includegraphics[width=.8\linewidth]{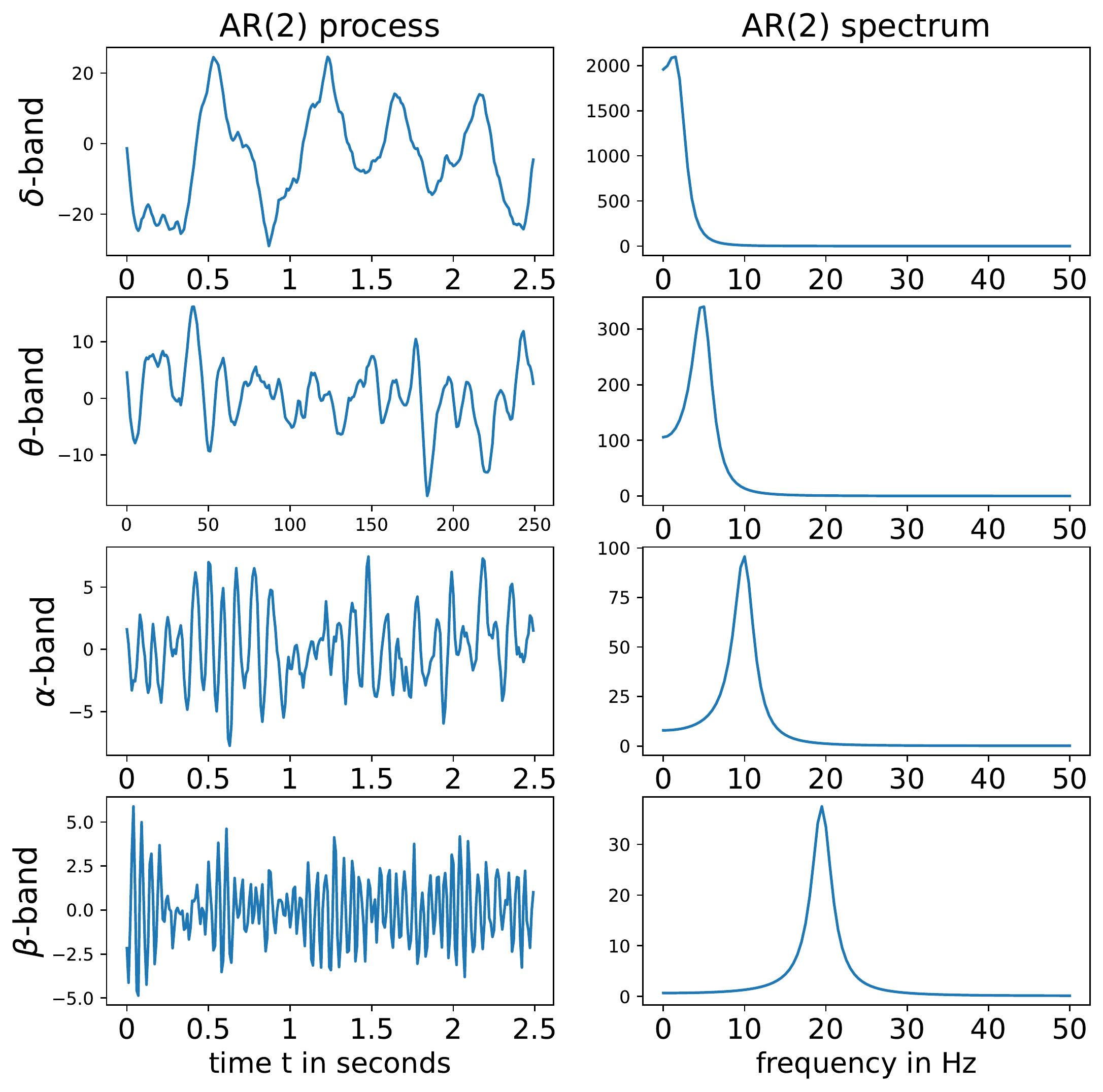}
    \caption{LEFT: AR(2) processes for different frequency bands. RIGHT: Corresponding spectrums. From top to bottom we have: delta-band with peak frequency at 2Hz, theta-band with peak frequency at 5Hz, alpha-band with peak frequency at 10Hz, alpha-band with peak frequency at 2Hz and beta-band with peak frequency at 19.5Hz with sampling rate (SR) of $100Hz$.} 
    \label{fig:AR2_time_series_and_spectrum}
\end{figure}

A linear mixture of second order autoregressive processes (AR(2) processes) can be used to simulate the brain oscillatory activity at specific frequency bands. An AR(2) process with a spectral peak at some frequency can be used to describe a latent process as follows:
\begin{align}
    Z(t) &= \phi_1 Z(t-1) + \phi_2 Z(t-2) + W(t) \label{Eq:AR2}
\end{align}
where $W(t)$ is white noise process with $\mathbf{E} \hspace{1mm} W(t) = 0$ and $Var \hspace{1mm} W(t) = \sigma^2$; the relationship between the AR(2) model parameters $\phi_1$ and $\phi_2$ and the spectral properties, namely the frequency peak and bandwidth, will be derived as follows. Note that Equation \ref{Eq:AR2} can be rewritten as $W(t) = (1 - \phi_1 B^1 - \phi_2 B^2)Z(t)$ where the operator $B^k Z(t) = Z(t-k)$ for $k=1, 2$. The AR(2) characteristic polynomial function is:
\begin{align}
    \Phi(r) = 1 - \phi_1 r^1 - \phi_2 r^2.
\end{align}
Consider the case when the roots of the $\Phi(r)$, denoted by $r_1$ and $r_2$, are (non-real) complex-valued and hence can be expressed as $r_1 = M \exp(i 2 \pi \psi)$ and $r_2 = M \exp(-i 2 \pi \psi)$ where the phase $\psi \in (0, 0.5)$ and the magnitude $M>1$ to satisfy causality, \cite{TSA_SHUMWAY_STOFFER}. For this latent process $Z(t)$, suppose that the sampling rate is denoted by $SR$ and the peak frequency is $f \in$($f_{\min}$;$f_{\max}$). Then the roots of the AR(2) latent process are $r_1 = M \exp(i 2 \pi \psi)$ and $r_2 = M \exp(-i 2 \pi \psi)$ where the phase $\psi = f / SR$. In practice, if the sampling rate $SR$ is 100 Hz and we wish to simulate an alpha-band latent process where the peak if at 10 Hz, then it is necessary to set $\psi = 10 / 100$ and the root magnitude $M$ to some number greater than 1 but "close" to 1 so that the spectrum of $Z(t)$ is mostly concentrated on the frequency band $f$-Hz. The corresponding AR(2) coefficients are derived to be $\phi_1 = \frac{2 \cos (2 \pi \psi)}{M}$ and $\phi_2 = -\frac{1}{M^2}$. Some examples of such stationary AR(2) processes as well as their corresponding spectrum is given in Figure \ref{fig:AR2_time_series_and_spectrum}.

\subsection{Mixtures of AR(2) processes}

In TDA applications, Rips-Vietoris filtrations are often applied to multivariate time series data (\cite{TDA_TSA_1}, \cite{TDA_TSA_2} and \cite{TDA_MULTIVARIATE_TS_ANASS}). These filtrations are often constructed from clouds of points, or from from a weighted network. Traditionally, due to their stochastic nature brain signals have often been modeled using their underlying dependence networks. For instance in \cite{GRAPHS_BRAIN_DATA}, the authors use graph theoretical methods on complex brain networks. 

To replicate a specific dependency pattern in the dependence network of a multivariate time series, it is necessary to emulate the decay in the dependence structure as time series components get farther away from each other. First sample a graph $G = (N, E)$ with the desired structure (i.e., cycles or holes etc.), second define the observed time series components as mixtures of the latent processes, such that components near to each other in the graph share more latent processes, which makes them more dependent on one another, while components far away in the graph will share fewer latent components, resulting in lower interdependence. Let $Z_p(t)$ be the latent iid AR(2) processes centered around a specific frequency band. Therefore, to generate multivariate time series with a desired dependence patterns (as defined by the graph G) the following model is suggested:
\begin{align}
    Y_p(t) &= \sum_{j=1}^P W_{p, q} Z_q(t) + \epsilon_q(t) \label{Eq:mixture_model} \\
    W_{p, q} &= 
        \begin{cases}
        \frac{1}{1 + d_G(p, q)}, &\text{ if } d_G(p, q) \leq K,\\
        0, &\text{ if } d_G(p, q) > K \label{Eq:mixing_weights}
        \end{cases}
\end{align}
We generate a $P$-dimensional vector of observations $Y(t) = [Y_1(t), ..., Y_P(t)]' \in \mathbb{R}^P$ that is a linear mixture of $P$ latent iid AR(2) processes $Z_1(t), ..., Z_P(t)$ according to Equation \ref{Eq:mixture_model},  whith $\mathbb{E} \big(Z_p(t)\big)=0$, and $Var\big(Z_p(t)\big)=1$. The mixing weights $W_{p, q}$ contain the information about the importance of the $q$-th latent AR(2) process $Z_q(t)$ in the $p$-th observed component $Y_p(t)$, and as defined by Equation \ref{Eq:mixing_weights} the weights are chose to be inversely proportional to the distance $d_G(p,q)$ between the nodes in the graph, and $K$ being the maximum distance threshold that is considered, in practice we take $K=2$ or $K=3$. Theoretically, any distance-decreasing function might be used. However, selecting a faster decay (such as exponential decay) could result in a too-sharp decrease in the dependence based distance, making it more challenging to identify the topological features in the filtration.

Let the $P$-dimensional observed vector to be $Y(t) = [Y_1(t), \hdots, Y_P(t)]'$. Then we have the following Cramer representation:
\begin{align}
    Y(t) &= \int_{-1/2}^{1/2} \exp(i2\pi \omega t) dX(\omega),
\end{align}
where the $X(\omega)$ is a $P$-variate random process whose mean is zero with orthogonal increments having the following covariance:
\begin{align}
    Cov(dX(\omega), dX(\lambda)) &=         
        \begin{cases}
            f(\omega) d\omega d\lambda \text{ if } \omega = \lambda + 2\pi k, \text{ k an integer}, \\
            0,  \text{ otherwise.}
        \end{cases}
\end{align}
and $f(\omega)$ is the spectral density matrix. If we define the filtered components at band $\Omega$ to be: 
\begin{align}
    Y_{1, \Omega}(t) &= \sum_\ell \Psi_\ell Y_1(t-\ell) \\ 
     & \vdots \notag \\
    Y_{P, \Omega}(t) &= \sum_\ell \Psi_\ell Y_P(t-\ell)
\end{align}
where the filter $\Psi_\ell$ is the band pass filter centered around frequency band $\Omega$.  In observation, coherence between $Y_1(.)$ and $Y_2(.)$ at frequency band $\Omega$ is just the squared correlation between $Y_{1, \Omega}(t)$ and $Y_{2, \Omega}(t)$, as explained in \cite{ONBAO_BELLEGEM} with more details and rigor. Coherence will then be used (via a decreasing transformation) to define frequency-specific distance between time series components $Y_p(.)$ and $Y_q(.)$.

The spectral matrix $f(\omega)$ can be estimated parametrically (e.g., by fitting a VAR model), non-parametrically (by smoothing the periodogram) or semi-parametrically (\cite{FUNCTIONAL_CON_1} \cite{FUNCTIONAL_CON_2}). In our case we will be using the smoothed periodogram approach. The Fourier $P$-dimensional coefficient at frequency $\omega_k$ are defined as:
\begin{align}
    d(\omega_k) &= \frac{1}{\sqrt{T}} \sum_{t=1}^T Y(t)\exp{(-i\omega_k t)},  \label{Eq:fourier_coef}
\end{align}
then the Fourier periodogram is defined to be:
\begin{align}
    I(\omega_k) &= d(\omega_k){d(\omega_k)}^*, \label{Eq:periodogram}
\end{align}
with $I(\omega_k)$ being a $P\times P$ matrix, as a consequence of this definition it can be shown that the periodogram $I(\omega)$ is unbiased. However, it is not a consistent estimator of the spectral matrix as the asymptotic variance does not decrease to zero when we get more and more observations. Hence, we construct a mean-squared consistent estimator to be:
\begin{align}
    \widehat{f}(\omega_k) &= \sum_\omega k_h(\omega - \omega_k) I(\omega) \label{Eq:smoothed_periodogram}
\end{align}
where $k_h(\omega - \omega_k)$ is a smoothing kernel centered around $\omega_k$ and $h$ is the bandwidth parameter. In order to define our distance function, first we define coherence as follows:
\begin{align}
    \mathcal{C}\Big(Y_{p}(.), Y_{q}(.), \omega\Big) &= \frac{|\widehat{f}_{p, q}(\omega)|^2}{\widehat{f}_{p, p}(\omega)\widehat{f}_{q, q}(\omega)} \in [0, 1], \label{eq:coherence}
\end{align}
then we define the dependence-based frequency-specific distance function to be a decreasing function (e.g., $\mathcal{G}(x) = 1-x$) of coherence:
\begin{align}
    \mathcal{D}\Big(Y_{p}(.), Y_{q}(.), \omega\Big) &= \mathcal{G}\Big(\mathcal{C}(Y_{p}(.), Y_{q}(.), \omega)\Big). \label{eq:dependence_based_distance}
\end{align}

In the following, using the ideas explained previously we start by generating multivariate time series data with specified dependence patterns. We explain how dependence information contained in the graph $G$ can be encoded in the homology structure of connectivity network, using a first example with one main cycle then a second example with two main cycles in the dependence network.

\section{Multivariate time series with cyclic patterns}

In this section, the goal is to develop a procedure for simulating multivariate time series with specific complex dependence structures. This is essential for evaluating TDA methods through simulations and also for conducting statistical inference in TDA. The topology of the brain network is known to be organized according to principles that maximize the flow of information and minimize the energy cost for maintaining the entire network (\cite{STRUCTURE_FUNCTION_BRAIN_NETWORKS}, \cite{BRAIN_NETWORKS_ORGANIZATION}, \cite{SMALL_WORLD_BRAIN} and \cite{SMALL_WORLD_NETWORKS}). However, some conditions may affect that organization by altering the structural connectivity of the brain (e.g., Alzheimer's disease) of by changing the functional connectivity (e.g., ADHD) (\cite{FUNCTIONAL_CONNECTIVITY_ALZHEIMER} and \cite{BRAIN_NETWORKS_HEALTHY_DISEASE}). Therefore, it is of interest to simulate multivariate time series with a given number of cycles.

\subsection{One main cycle pattern}
\label{subsec:one_main_cycle}

In this first example we generate a multivariate time series with one cycle in the dependence structure. We want time series components that are relatively close to each other to be more strongly dependent than components that are farther apart.
\begin{figure}[ht]
    \centering
    \includegraphics[width=.4\linewidth]{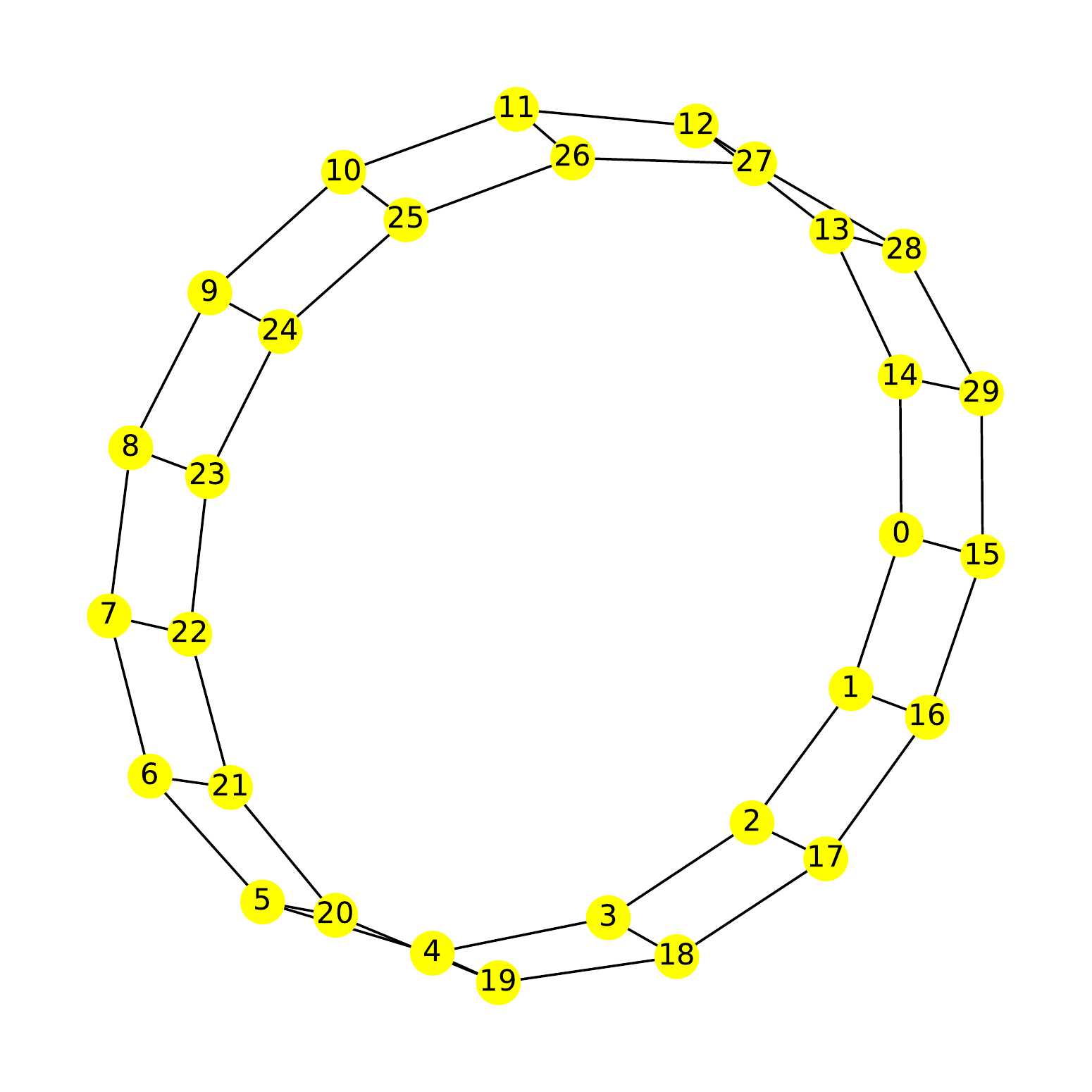}
    \caption{Cyclic dependence graph - Circular Ladder model.}
    \label{fig:one_cycle_dependence_pattern}
\end{figure}
Given the previous network definition of the circular ladder model, as displayed in Figure \ref{fig:one_cycle_dependence_pattern}, we can write down the expression for $K=2$, using Equation \ref{Eq:mixture_model} for the observed time series $Y(t) = [Y_{1}(t), \hdots, Y_{30}(t)]'$ and latent process $Z(t) = [Z_{1}(t), \hdots, Z_{30}(t)]'$ as follows:
\begin{align}
    Y(t) &= W Z(t) + \epsilon(t),
\end{align}
where $W_{p, q}$ is the contribution of the latent process $Z_q(t)$ in the observed process $Y_p(t)$ and is equal to $\frac{1}{1 + d_G(p, q)}$ if the distance $d_G(p, q)$ between nodes $p$ and $q$ is less than or equal to $2$ and $0$ otherwise. Thus, $W$ is a $30\times 30$ weight matrix, $Z(t)$ is a $30\times 1$ latent process vector and $\epsilon(t)$ is a $30\times 1$ noise vector. Using this circular ladder model we can generate and visualize the multivariate time series data as follows in Figure \ref{fig:TS_One_Cycle}:
\begin{figure}[ht]
    \centering
    \includegraphics[width=.7\linewidth]{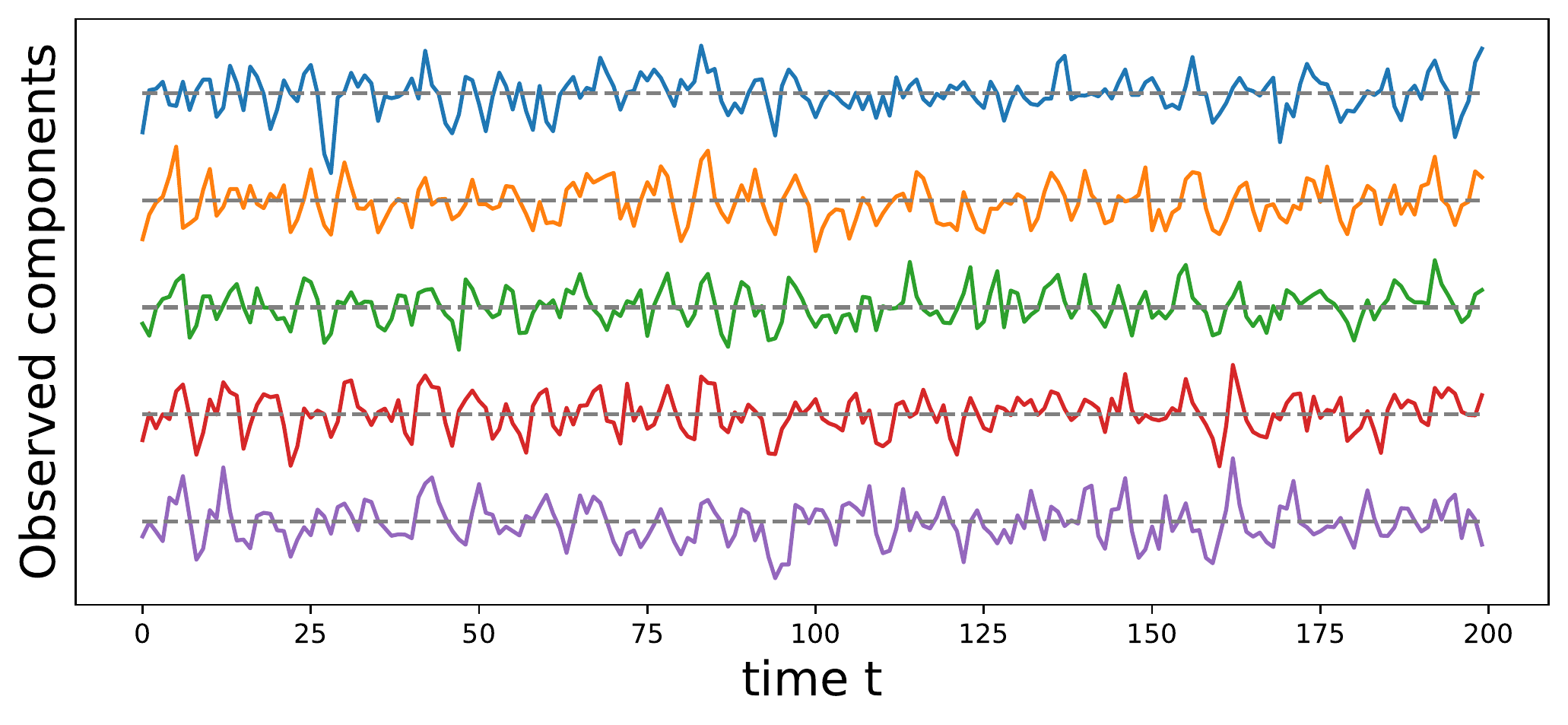}
    \caption{Generated time series with $P=30$. The following components are plotted from top to bottom:  $Y_0(t), Y_1(t), Y_2(t), Y_3(t), Y_4(t), Y_5(t)$.}
    \label{fig:TS_One_Cycle}
\end{figure}

Given the time series components formulation above. It is possible to compute the correlation, for example, for the pairs $Y_1(.)$ - $Y_0(.)$ and $Y_1(.)$ - $Y_2(.)$ and $Y_1(.)$ - $Y_6(.)$ this leads to $\mathcal{C}orr\Big(Y_{1}(.), Y_{0}(.)\Big) = \mathcal{C}orr\Big(Y_{1}(.), Y_{2}(.)\Big)$ which simplifies to $\frac{25/9}{1369/144} \sim 0.29$ and $\mathcal{C}orr\Big(Y_{1}(.), Y_{0}(.)\Big) = 0$. Therefore, the distance between components 1-0 and 1-2 is the same $0.69$, this is due to symmetry in the graph in Figure \ref{fig:one_cycle_dependence_pattern} and distance between component 1 and 6 is $1$ because they do not share any latent components. Therefore, the time series components 1 and 6 are farther apart (based on the dependence distance) than 1 and 0 or 1 and 2, which is exactly the desired property.

Now that we have a good intuition behind the mechanism that generates the time series components we can directly compute the coherence matrices for various frequency bands and analyze the topological patterns present in the resulting network.
\begin{figure}[ht]
    \centering
    \includegraphics[width=.6\linewidth]{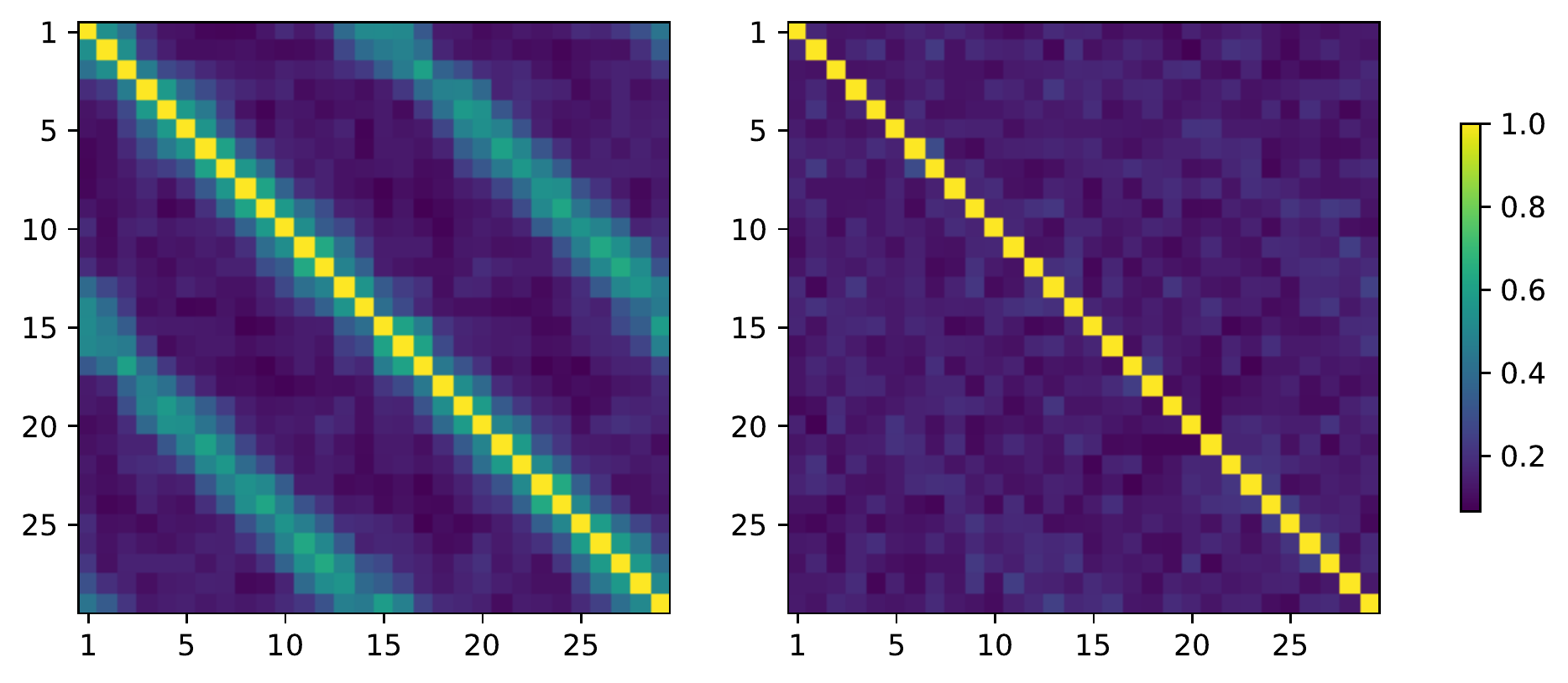}
    \caption{Estimated coherence for one cycle model. LEFT: Middle frequency band; RIGHT: High frequency band.}
    \label{fig:Coherence_One_Cycle}
\end{figure}
After computing the coherence matrices and therefore the distance matrices, it is now possible to build the Rips-Vietoris filtration and visualize the results in the persistence diagram form as follows:
\begin{figure}[ht]
    \centering
    \includegraphics[width=.6\linewidth]{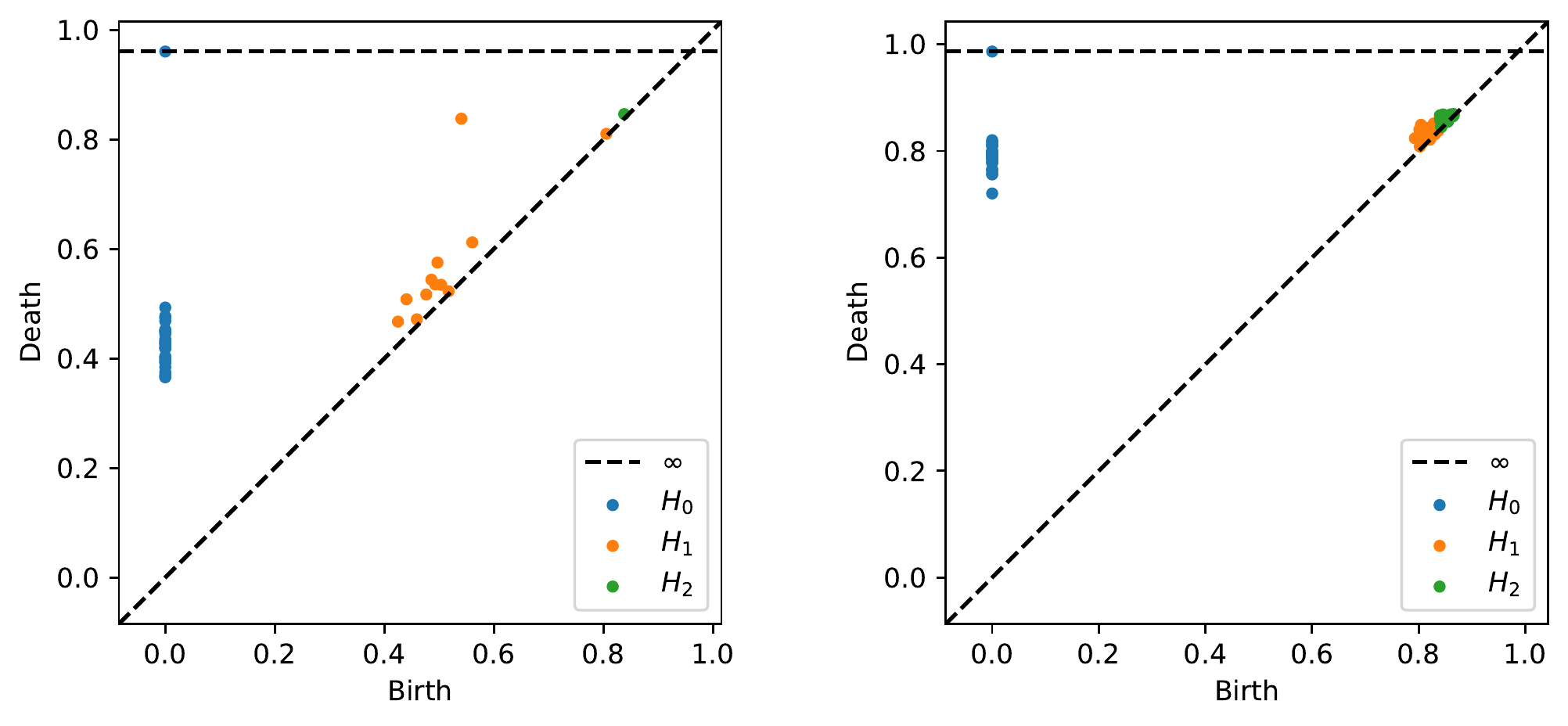}
    \caption{Estimated persistence diagram for one cycle model. LEFT: Middle frequency band; RIGHT: High frequency band.}
    \label{fig:PD_One_Cycle}
\end{figure}
The orange point in Figure \ref{fig:PD_One_Cycle} far from the diagonal represents the main cycle in the dependence structure. Whereas the orange dots near the diagonal represent the secondary cycle that are present all around the network, as can be seen in Figure \ref{fig:one_cycle_dependence_pattern}.

\subsection{Two main cycles pattern}
\label{subsec:two_main_cycle}

\begin{figure}[ht]
    \centering
    \includegraphics[width=.7\linewidth]{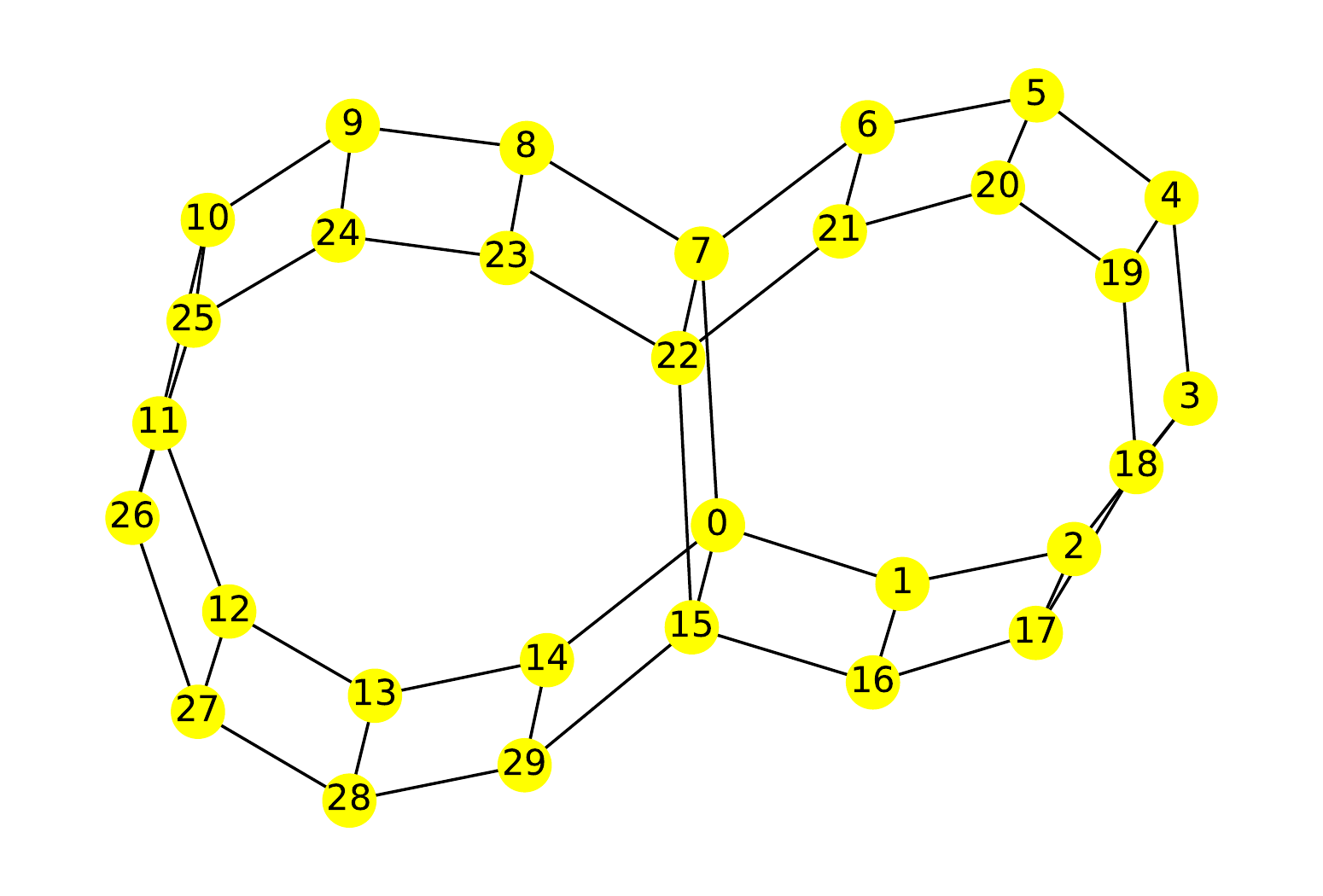}
    \caption{Cyclic dependence graph with two main cycles - Double circular ladder model. Every node corresponds to a time series component.}
    \label{fig:two_cycle_dependence_pattern}
\end{figure}

\begin{figure}[ht]
    \centering
    \includegraphics[width=.7\linewidth]{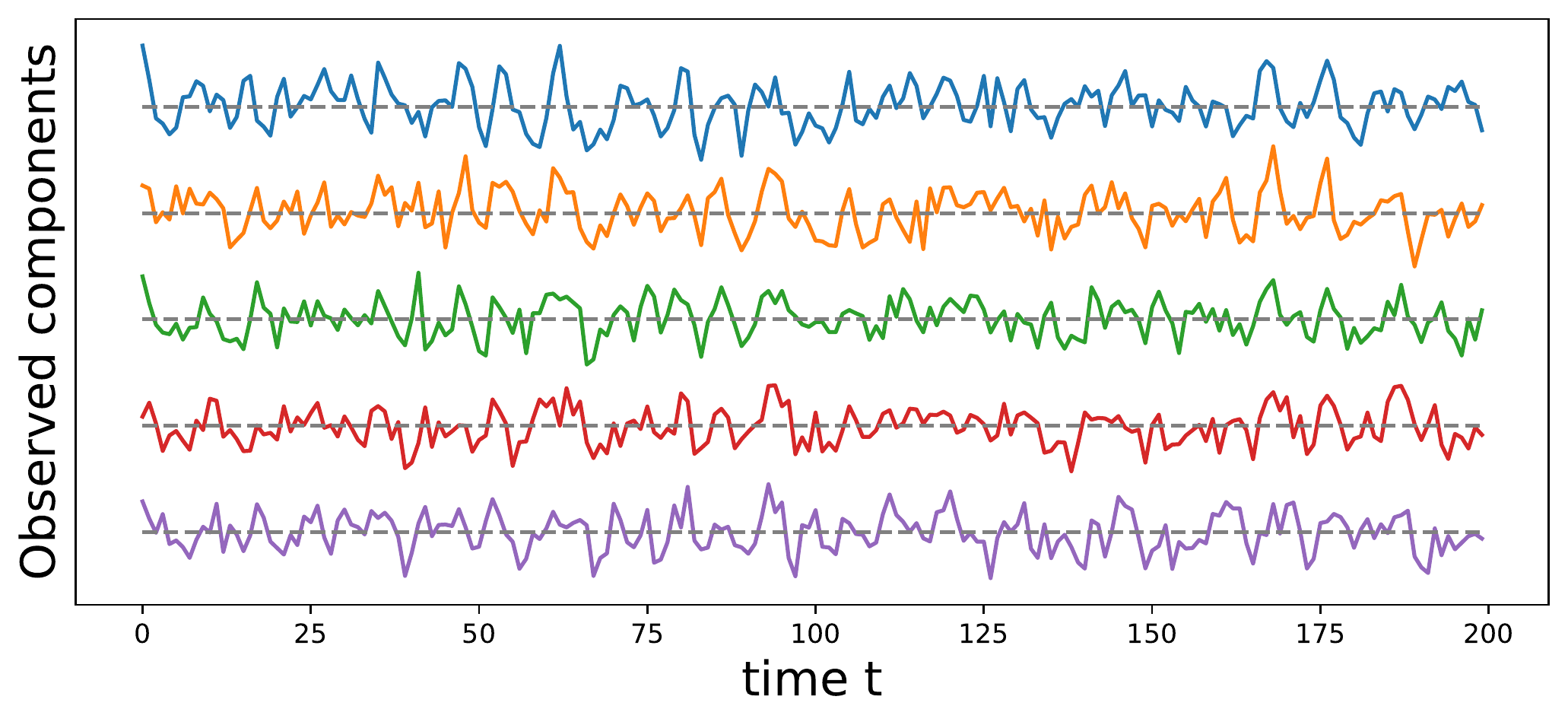}
    \caption{Generated time series with $P=30$. The following components are plotted from top to bottom:  $Y_0(t), Y_1(t), Y_2(t), Y_3(t), Y_4(t), Y_5(t)$, derived from the double circular ladder model.}
    \label{fig:TS_Two_Cycle}
\end{figure}

Here, we develop a model for generating multivariate time series with two cycles in the dependence structure. Similarly, from Equation \ref{Eq:mixture_model}, one can generate the multivariate time series with the double circular ladder model as defined in Figure \ref{fig:two_cycle_dependence_pattern}. Using this mechanism we can generate and visualize the multivariate time series as follows, see Figure \ref{fig:TS_Two_Cycle}.



Without computing the coherence expressions, it can be noted that the coherence between any given pair of channels with a subnetwork will not necessarily vanish as the diameter of the subnetwork became much smaller than it was with the previous example from Section \ref{subsec:one_main_cycle}. Intuitively, components that are farther apart will have weaker dependence and components that are closer will exhibit stronger dependence as they share more latent processes. We can directly compute the coherence matrices for middle and high frequency bands and analyze the topological patterns present in the resulting network.
\begin{figure}[ht]
    \centering
    \includegraphics[width=.6\linewidth]{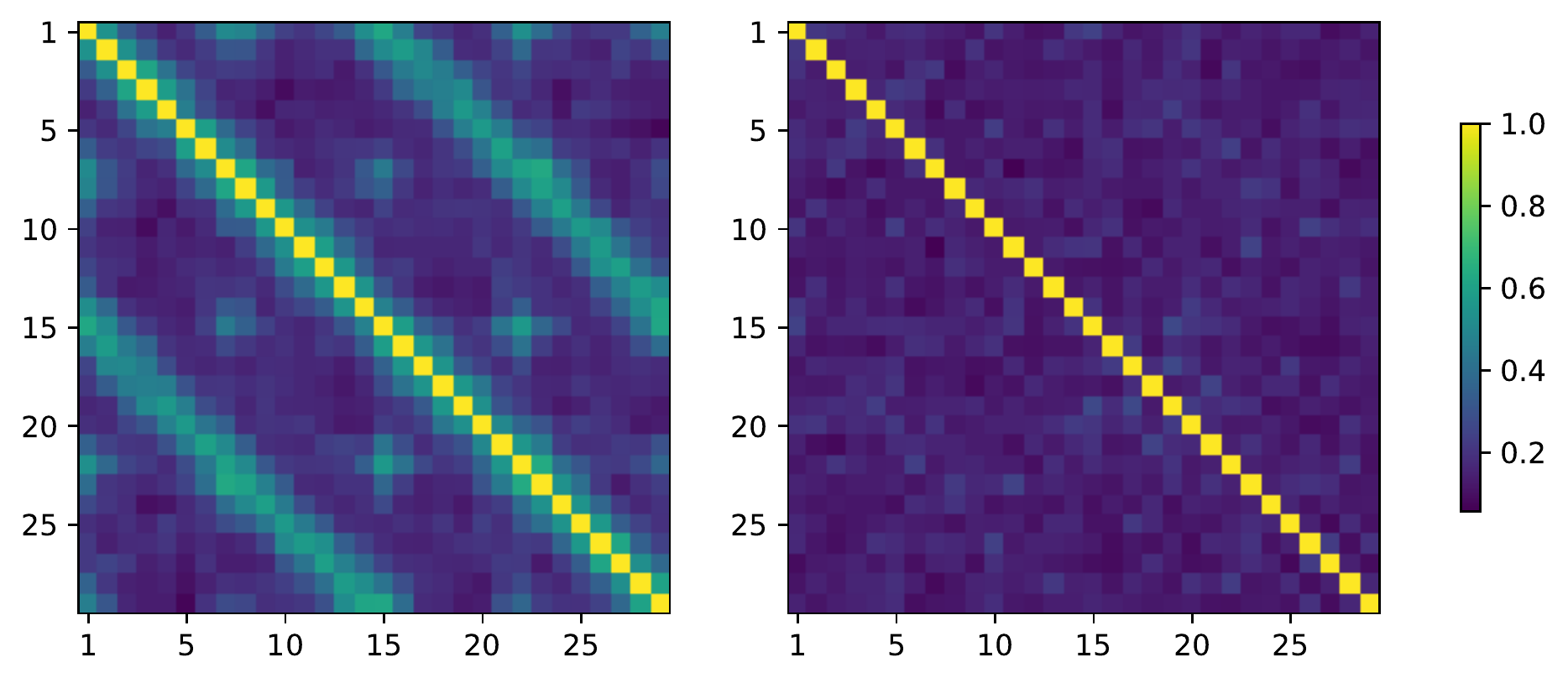}
    \caption{Estimated coherence from two cycles model. LEFT: Middle frequency band; RIGHT: High frequency band.}
    \label{fig:Coherence_Two_Cycles}
\end{figure}
After estimating the coherence matrices and therefore the distance matrices we can apply the tools of TDA, i.e., building the Rips-Vietoris filtration and visualize the results in the persistence diagram form as follows:
\begin{figure}[ht]
    \centering
    \includegraphics[width=.6\linewidth]{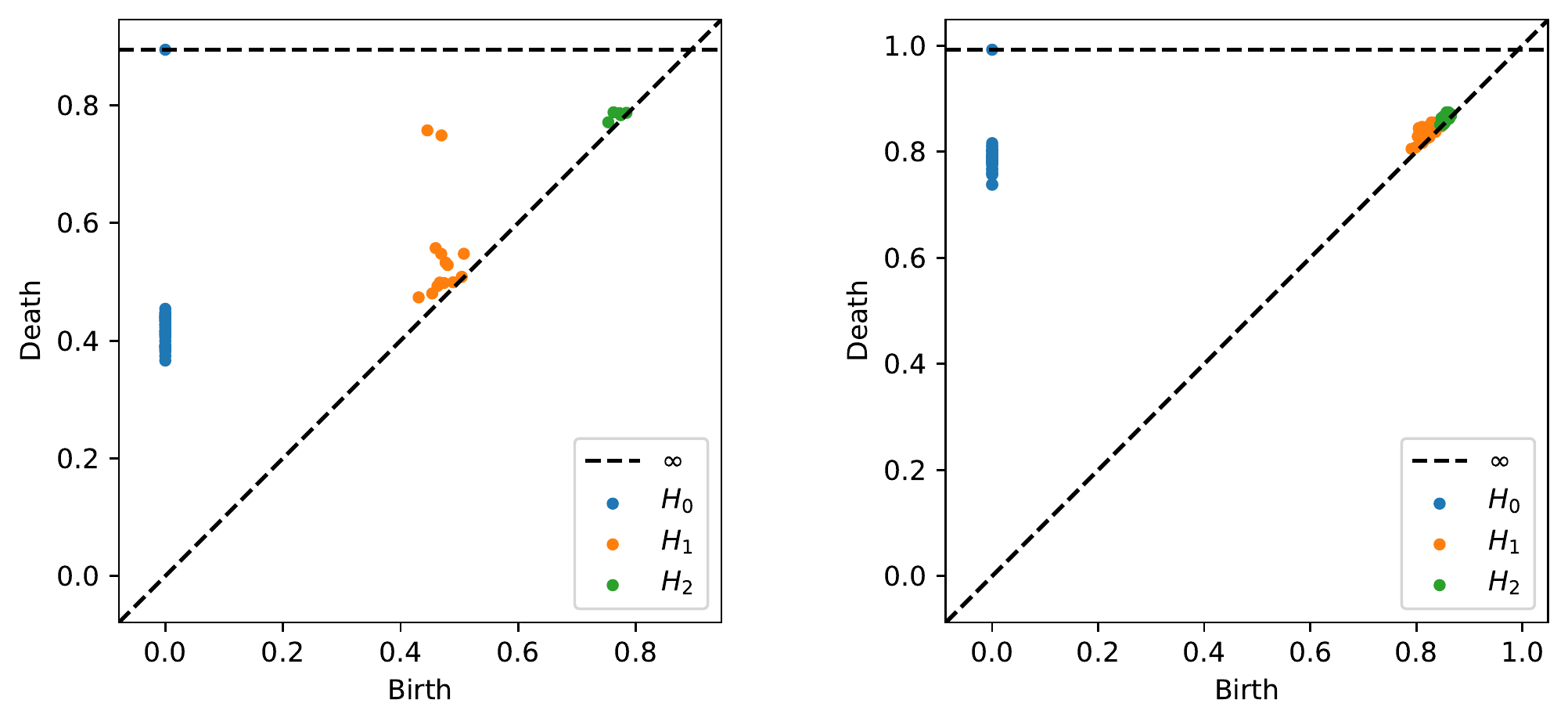}
    \caption{Estimated persistence diagram for two cycles model. LEFT: Middle frequency band; RIGHT: High frequency band.}
    \label{fig:PD_Two_Cycle}
\end{figure}
The x-axis represents the birth time, while the y-axis represents the death time, all the point representing valid features in the diagram have to lay above the diagonal line since the death time is larger than the birth time, i.e., $d_i>b_i$. The orange points far from the diagonal represents the two main cycles in the dependence structure. Whereas the orange dots near the diagonal represent the secondary cycle that are present all around the network, as can be seen in Figure \ref{fig:two_cycle_dependence_pattern}.

\section{Generating multivariate time series with general patterns in its dependence network}

\begin{figure}[ht]
    \centering
    \includegraphics[width=.8\linewidth]{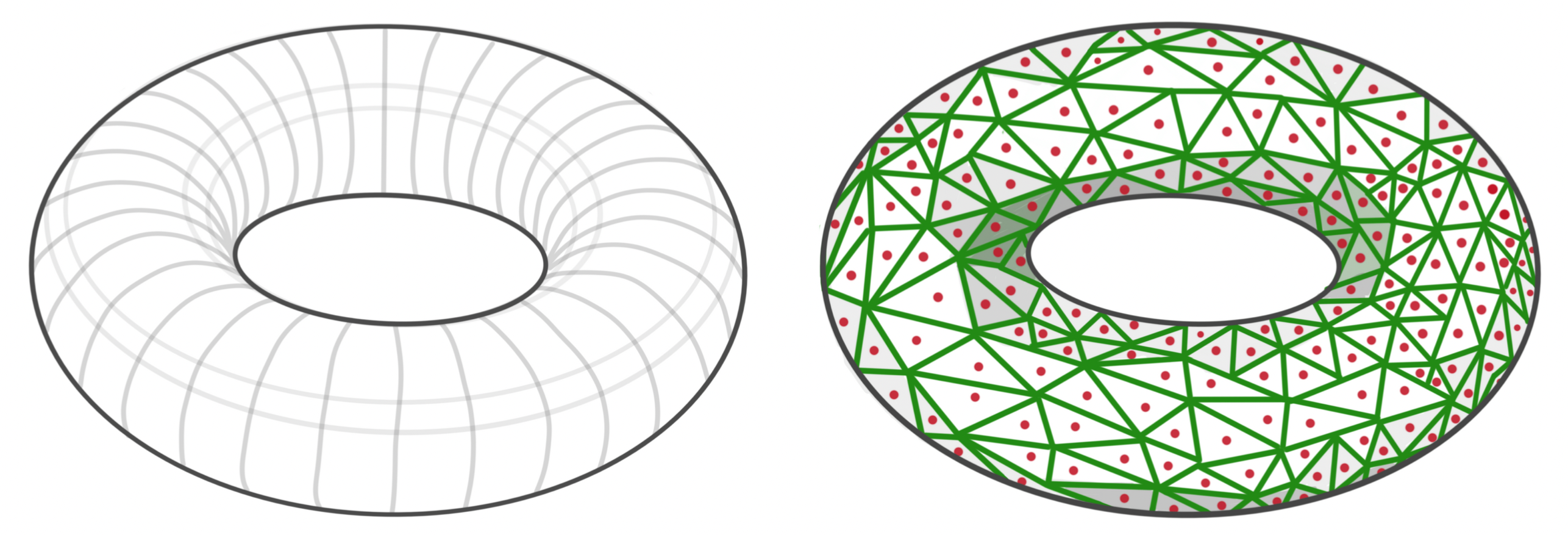}
    \caption{LEFT: Simple torus dependence pattern; RIGHT: Sampled dependence pattern.}
    \label{fig:simple_torus_dependence_pattern}
\end{figure}

\begin{figure}[ht]
    \centering
    \includegraphics[width=\linewidth]{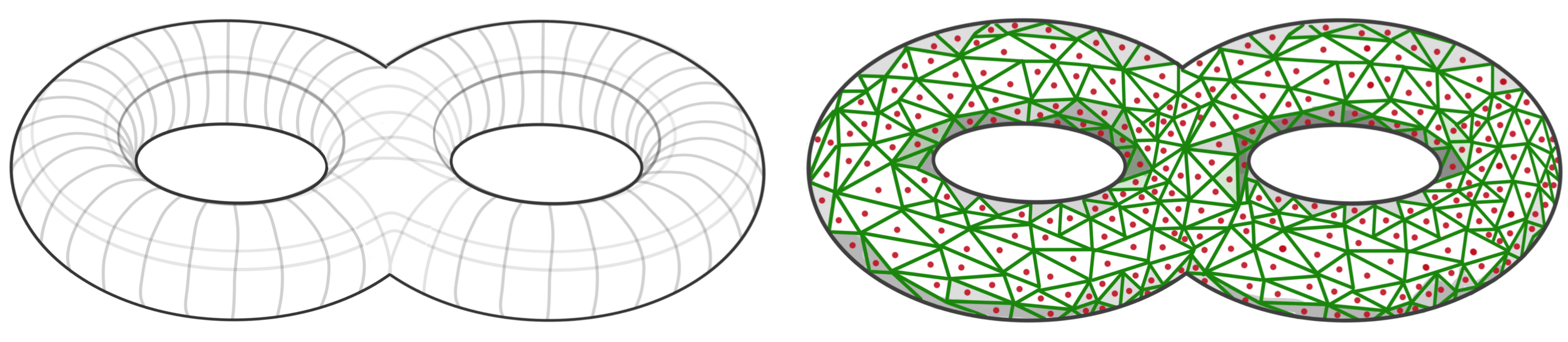}
    \caption{LEFT: Double torus dependence pattern; RIGHT: Sampled dependence pattern.}
    \label{fig:double_torus_dependence_pattern}
\end{figure}

Depending on the application of interest, the simulated patterns presented above may not be sufficient. However, the methodology is general and can be used to define many patterns in the dependence network of a multivariate time series. The goal in this section is to develop a  novel robust procedure for simulating multivariate time series with complex dependence structures. Suppose that the interest is on a specific connectivity pattern, such as a torus or a double torus, see respectively left hand side of Figures \ref{fig:simple_torus_dependence_pattern} and \ref{fig:double_torus_dependence_pattern}. 


\subsection{Defining the graph structure}

After defining the shape of the manifold of reference $\mathcal{M}$ for the dependence structure, we need to define a graph $G_\mathcal{M} = (N_\mathcal{M}, E_\mathcal{M})$ by sampling points $x_i$ from such a topological structure (i.e., $x_i \in N_\mathcal{M} \subset \mathcal{M}$), as can be seen in red dots in right hand side of Figures \ref{fig:simple_torus_dependence_pattern} and \ref{fig:double_torus_dependence_pattern}. It is necessary to properly define this graph, because it is needed to compute the mixing weights in Equation \ref{Eq:mixing_weights}, which are a (decaying) function of the graph distance between nodes. The right hand side of Figures \ref{fig:simple_torus_dependence_pattern} and \ref{fig:double_torus_dependence_pattern}, display a Voronoi tessellation (green lines) over the manifold. For every pair of nodes (region centers) that share a common border, we add an edge in the set of edges $E_\mathcal{M}$ in the graph $G_\mathcal{M}$. Once the graph is properly defined, we need to generate a latent AR(2) process for every node in the graph, then to generate a new multivariate time series we can simply mix the latent processes using the weighted approach described in Equation \ref{Eq:mixture_model}.

\subsection{Sampling points from a manifold}

In order to define the graph structure, it is necessary to have a mechanism to sample points uniformly from a manifold. Multiple sampling procedures have been proposed in the literature, for example \cite{MANIFOLD_SAMPLING_1}, \cite{MANIFOLD_SAMPLING_2}, \cite{MANIFOLD_SAMPLING_4} and \cite{MANIFOLD_SAMPLING_3}. It may be relatively straightforward to sample from simple manifolds, such as circles or spheres, because it is simple to parameterize the entire manifold, for instance using polar or spherical coordinates. However, generally speaking, sampling from more intricate manifolds can be rather difficult. This sampling problem (from a given manifold) is closely related to Bertrand's Paradox and the principle of indifference. Indeed, for such problem to display a unique solution one has to properly define the problem at hand and what is meant by sampling in a non-ambiguous way (\cite{BERTRAND_PARADOX_1}, \cite{BERTRAND_PARADOX_2}). For instance, considering the one dimensional circle embedded in $\mathbb{R}^2$, every point $p_1\in \mathcal{M}_1$ of the manifold $\mathcal{M}_1$ can be represented by an angle $\theta$:
\begin{align}
    \mathcal{M}_1 &= \{ (x, y) | x^2 + y^2 = r^2\}, \label{Eq:manifold_definition_circle}\\
    p_1 &= \big(r cos(\theta), r sin(\theta)\big), \hspace{3mm} \theta \in [0, 2\pi], \label{Eq:sphere_parametrization_circle}
\end{align}
Similarly, considering the two dimensional sphere embedded in $\mathbb{R}^3$, every point $p_2\in \mathcal{M}_2$ can be represented by a pair of coordinates:
\begin{align}
    \mathcal{M}_2 &= \{ (x, y, z) | x^2 + y^2 + z^2 = r^2\} \label{Eq:manifold_definition_sphere}\\
    p_2 &= \big(r sin(\theta)cos(\phi), r sin(\theta)sin(\phi), r cos(\theta)\big), \phi \in [0, 2\pi], \theta \in [-\pi/2, \pi/2] \label{Eq:sphere_parametrization_sphere}
\end{align}
For instance, the parameterizations in Equation \ref{Eq:sphere_parametrization_circle} correctly characterizes the circle. Hence, it is possible to sample points $p_i$ from the manifold $\mathcal{M}_1$ using the following procedure:
\begin{align}
    \theta_1 &\sim \mathcal{U}(0, 2\pi) \\
    p_1 &= \big(r cos(\theta_1), r sin(\theta_1)\big)
\end{align}
Similarly, the parameterization in Equation \ref{Eq:sphere_parametrization_sphere} correctly characterizes the two dimensional sphere of radius $r$. Hence, to sample points $p_i$ from $\mathcal{M}_2$ we can use the following procedure:
\begin{align}
    \theta_2 &\sim \mathcal{U}(0, 2\pi) \\
    \phi_2 &\sim \mathcal{U}(-\pi/2, \pi/2) \\
    p_2 &= \big(r sin(\theta_2)cos(\phi_2), r sin(\theta_2)sin(\phi_2), r cos(\theta_2)\big)
\end{align}
Both examples presented above rely on parameterized immersions. When the chosen parameterization $f : S \to \mathcal{M}$  is not volume-preserving, the resulting sample will not be uniform. Indeed, this approach will lead to compressed regions being oversampled and expanded regions being undersampled, i.e., based on uniform sampling in the parameter space $S$ the sampled points in $\mathcal{M}$ are denser in regions where the parameterization $f$ has higher curvature, \cite{MANIFOLD_SAMPLING_1} for more details. For instance, in the first example the sample is uniform, however, in the second example the sample will not be uniform as there are compressed regions around the poles and expanded regions farther away from the poles. To remedy this issue, \cite{MANIFOLD_SAMPLING_1} suggest an approach that consists in generating a large sample using the previous approach then discarding some of the samples to correct for the compressed and expanded regions. The rejection rate is chosen as a function of the determinant of the Jacobian of the parameterization $f : S \to \mathcal{M}$. Other interesting approaches have been proposed in the literature, such as the ones in \cite{MANIFOLD_SAMPLING_2} and \cite{MANIFOLD_SAMPLING_3}, however, these approaches provides tools for sampling for simple manifolds.

When the manifold of interest is not very simple, such as a double torus in Figure \ref{fig:double_torus_dependence_pattern} or even more complicated surfaces, it can be quite challenging to generate a sample using the above mentioned approach, since in some cases there may not be a global parameterization. Indeed, for smooth manifolds the parameterization is only guaranteed locally, to parameterize the entire manifold it is necessary to look at what is known as an Atlas representation of the manifold, refer to \cite{MANIFOLD_BOOK} for more details regarding the parameterization of manifolds.

\begin{figure}[ht]
    \centering
    \includegraphics[width=.6\linewidth]{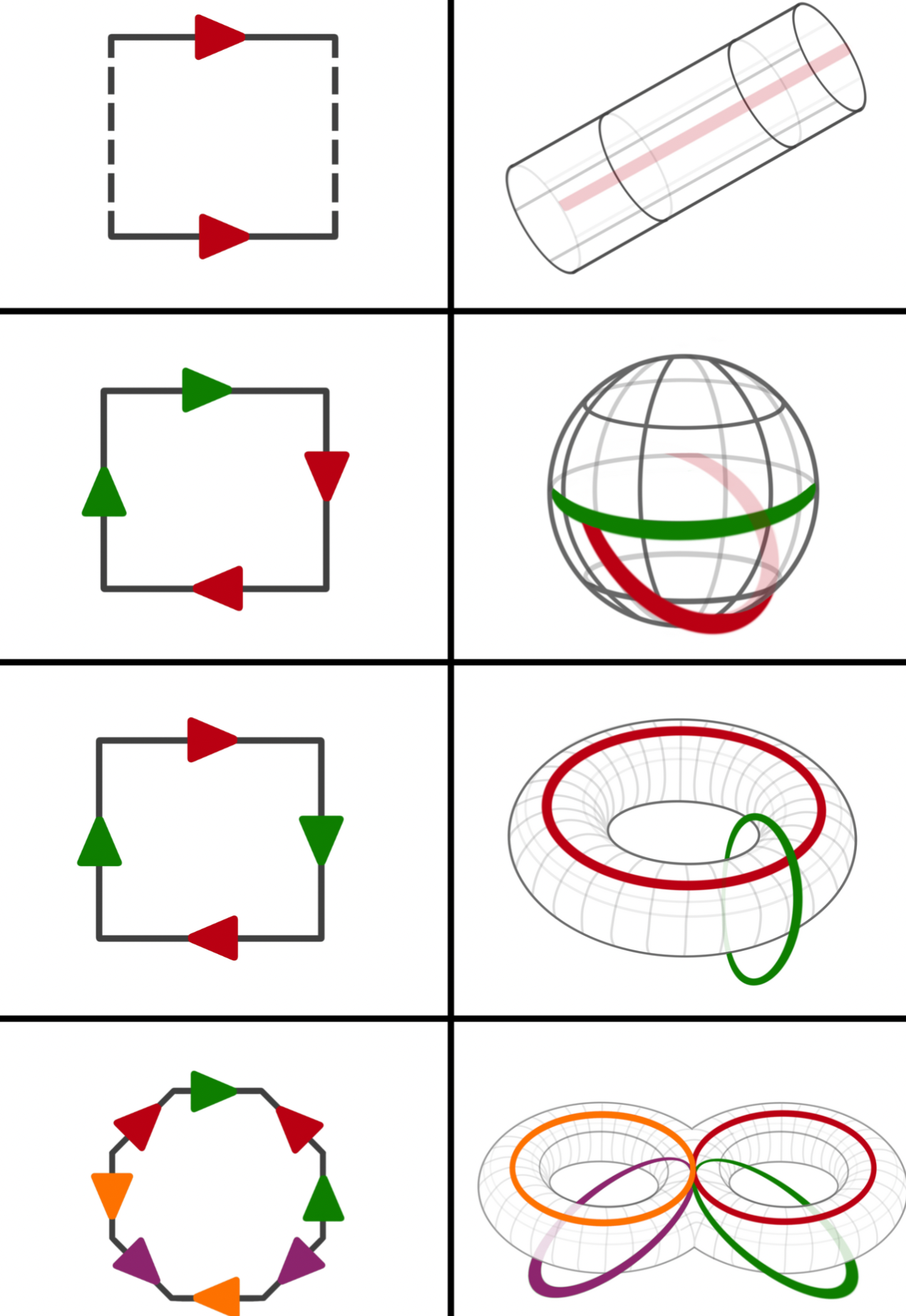}
    \caption{LEFT: Polygonal quotient group representation for a cylinder, sphere, torus and a double torus. RIGHT: Corresponding 3D visualization.}
    \label{fig:quotient_group_representation}
\end{figure}

\begin{figure}[ht]
    \centering
    \includegraphics[width=.7\linewidth]{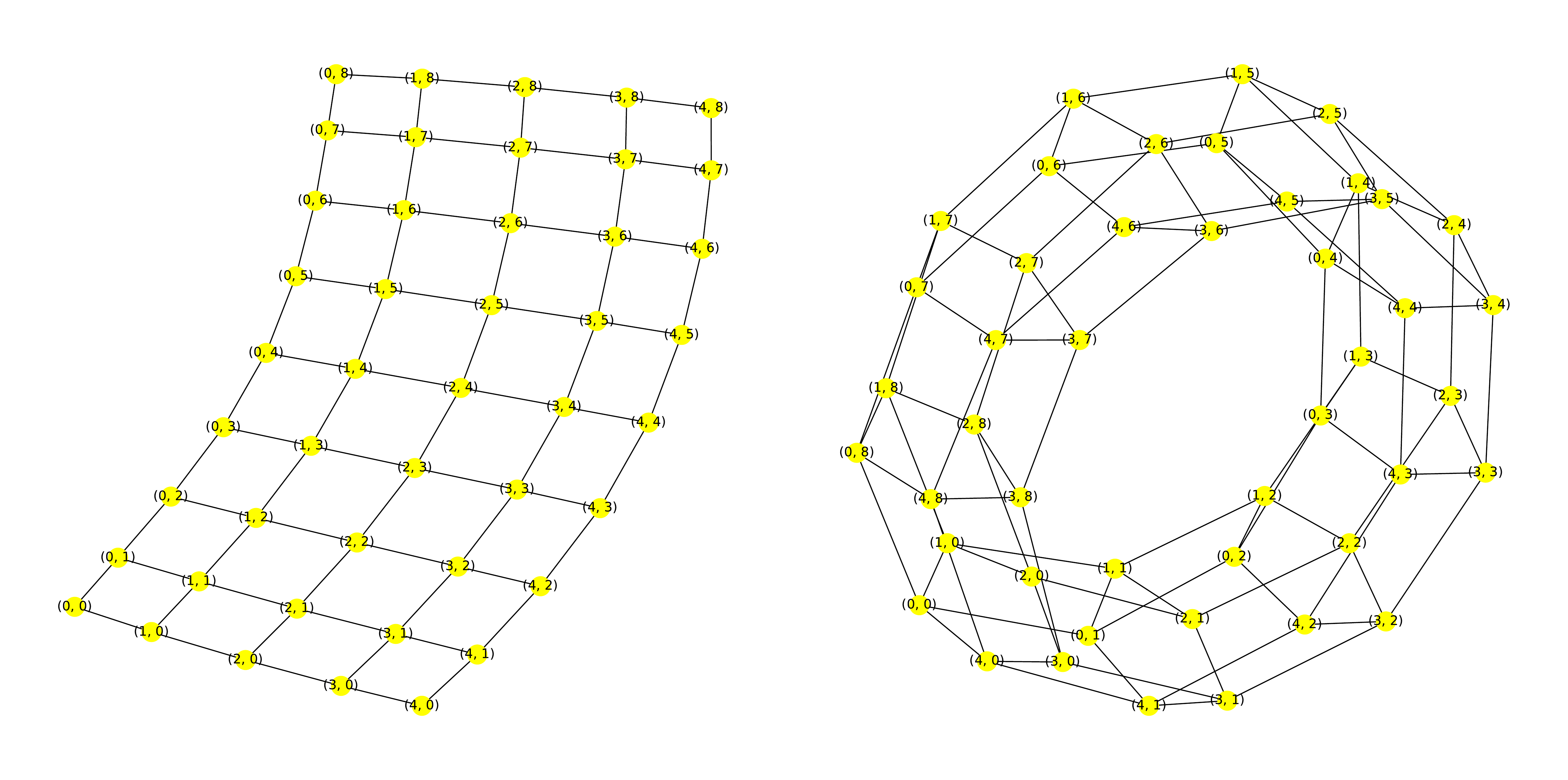}
    \caption{Torus graph pattern based on polygonal quotient group representation. LEFT: Regular sample from a rectangle; RIGHT: Torus graph after node identification.}
    \label{fig:quotient_group_sample_torus}
\end{figure}

\begin{figure}[ht]
    \centering
    \includegraphics[width=.7\linewidth]{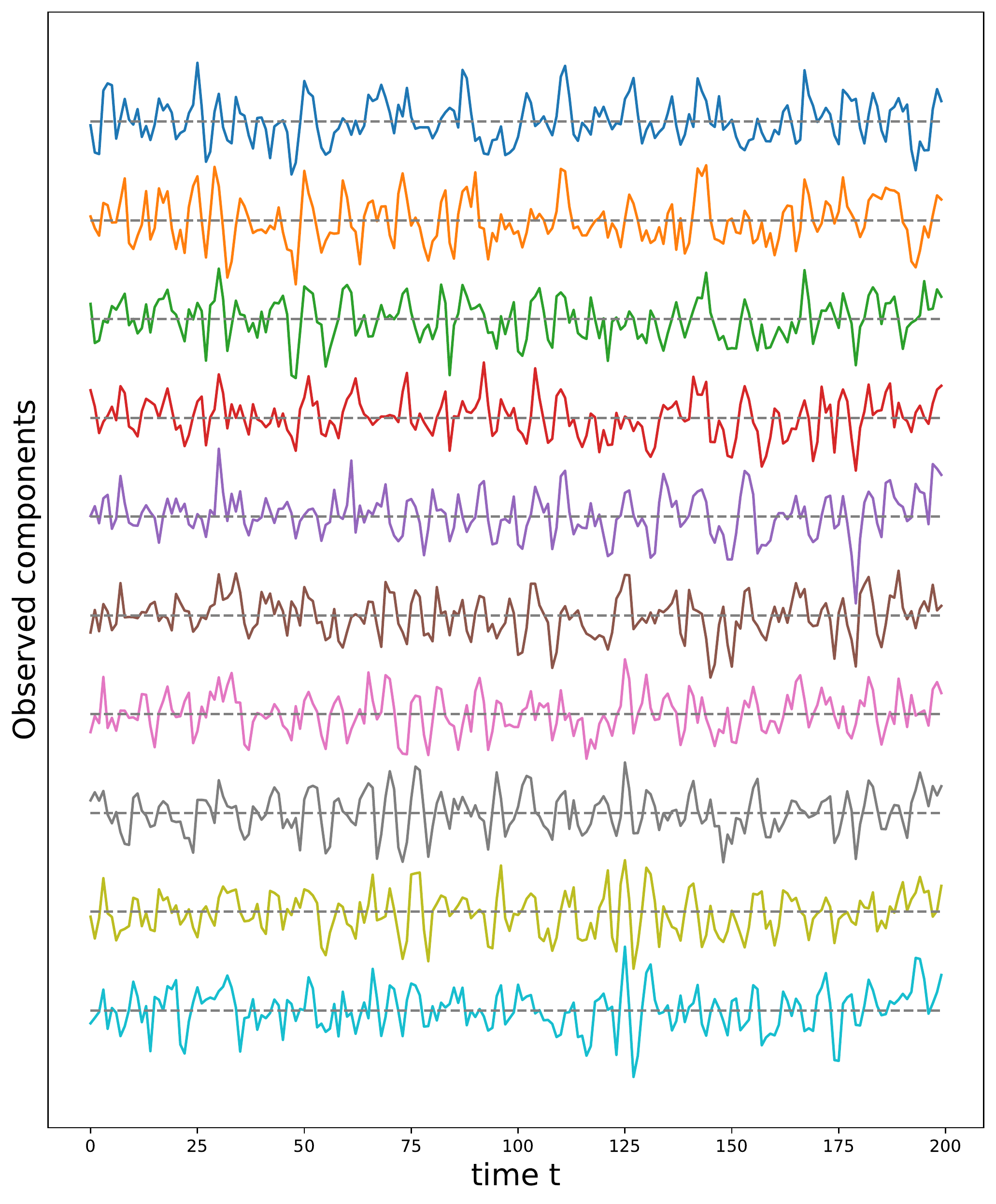}
    \caption{First ten time series components from a multivariate time series model generated using a torus structure with an initial grid of 9 by 17 nodes (i.e., $P=153$).}
    \label{fig:quotient_group_sample_torus_time_series}
\end{figure}

\begin{figure}[ht]
    \centering
    \includegraphics[width=.7\linewidth]{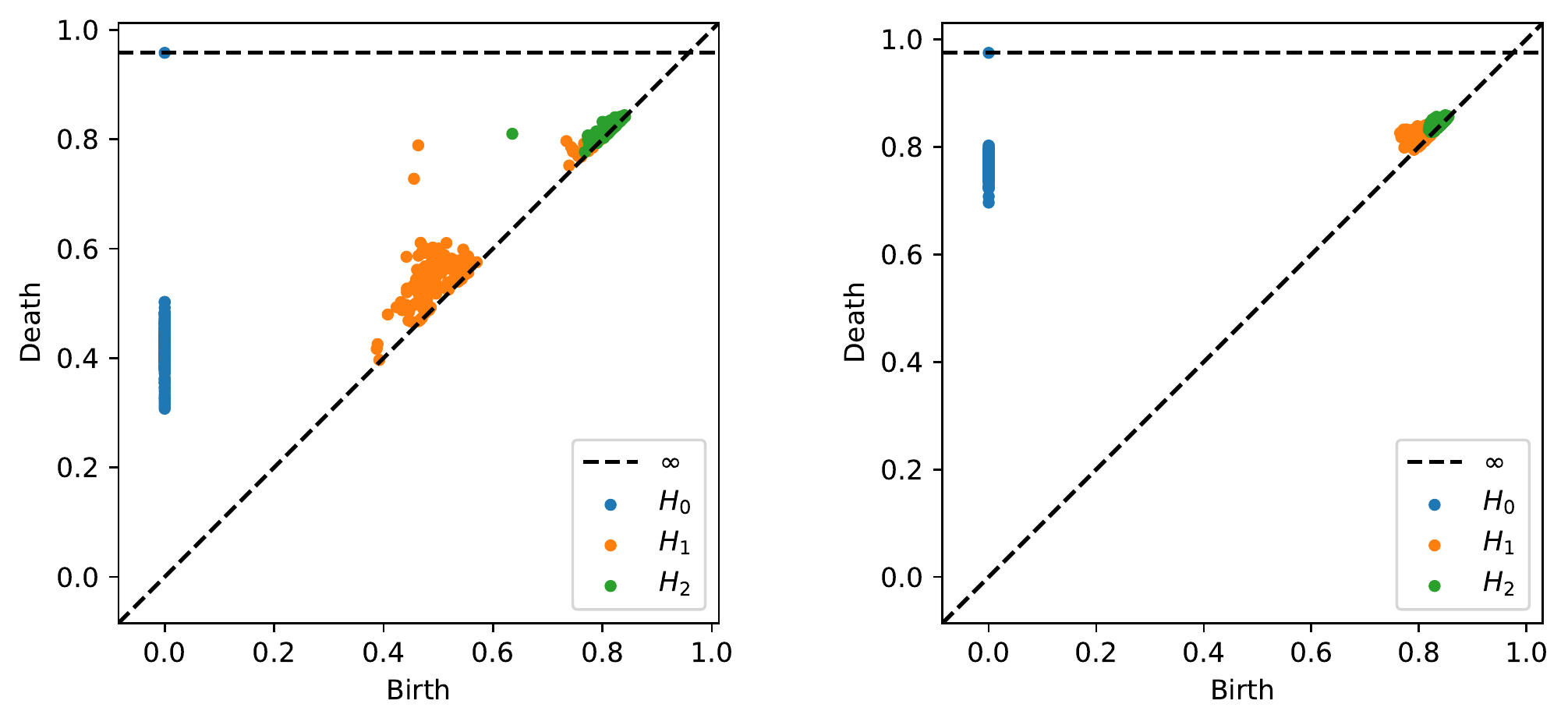}
    \caption{Persistence diagram based on a multivariate time series generated using a torus structure with an initial grid of 9 by 17 nodes,i.e., $P=153$, and $K=3$.}
    \label{fig:quotient_group_sample_torus_persistence_diagram}
\end{figure}

For these reasons, we suggest the following method to sample from a certain set of two-dimensional manifolds (surfaces), such as the sphere, torus, and double torus, which is based on the representation of these manifolds using quotient space of polygons, see Figure \ref{fig:quotient_group_representation}.
The interest behind this representation lies in the simplicity with which we can sample from the corresponding manifold. Indeed, once a polygonal representation at hand, one can sample uniformly from the flat polygons, then identify the nodes present on equivalent edges. In Figure \ref{fig:quotient_group_sample_torus}, it is possible to see both the initial sample form a rectangle and the constructed graph representing the torus manifold after node identification.
Using the same approach as described in Equation \ref{Eq:mixture_model}, we generate the torus multivariate time series, as can be seen in Figure \ref{fig:quotient_group_sample_torus_time_series}.
After estimating the coherence matrix for this multivariate time series at middle and high frequency bands we compute the persistence diagrams and find the following results, as displayed in Figure \ref{fig:quotient_group_sample_torus_persistence_diagram}.
This figure clearly shows the topological features of the targeted torus structure. Indeed, the two off-diagonal orange dots represent the two one dimensional wholes in a torus (circles surrounding each of the wholes) and one off-diagonal green dot representing the two-dimensional whole (or cavity inside of the torus), indeed it is known that the first three Betti numbers of a torus are $\beta_0 = 1$, $\beta_1 = 2$, $\beta_2 = 1$. 

The number of points to sample from the manifold depends on the type of topological feature we want to be able to detect. Generally speaking, for features to be detectable in the persistence diagram, the diameter of every subgraph surrounding the topological feature of interest needs to be at least of the same  magnitude or larger than twice the constant $K$ in the mixing equation. This is an important point to keep in mind. For instance, in Figure \ref{subsec:one_main_cycle} to be able to detect the main cycle we need the diameter ($P/4$) to be larger than $2K$, if $K=2$ then we need to chose $P\geq16$, if $K=3$ then we need to chose $P\geq24$ etc. In Figure \ref{subsec:two_main_cycle} to be able to detect the main cycles we need the diameter of the smallest subgraph surrounding one of the main cycles (roughly $P/8$ if both cycles are of comparable size) to be larger than $2K$, i.e., if $K=2$ then we need to chose at least $P\sim 32$. For this reason, we can detect only the main cycles and the secondary cycles appear like noise in the persistence diagrams, see Figure \ref{fig:PD_One_Cycle} and \ref{fig:PD_Two_Cycle}.

\section{Sensitivity to noise}

\begin{figure}[b]
    \centering
    \includegraphics[width=.8\linewidth]{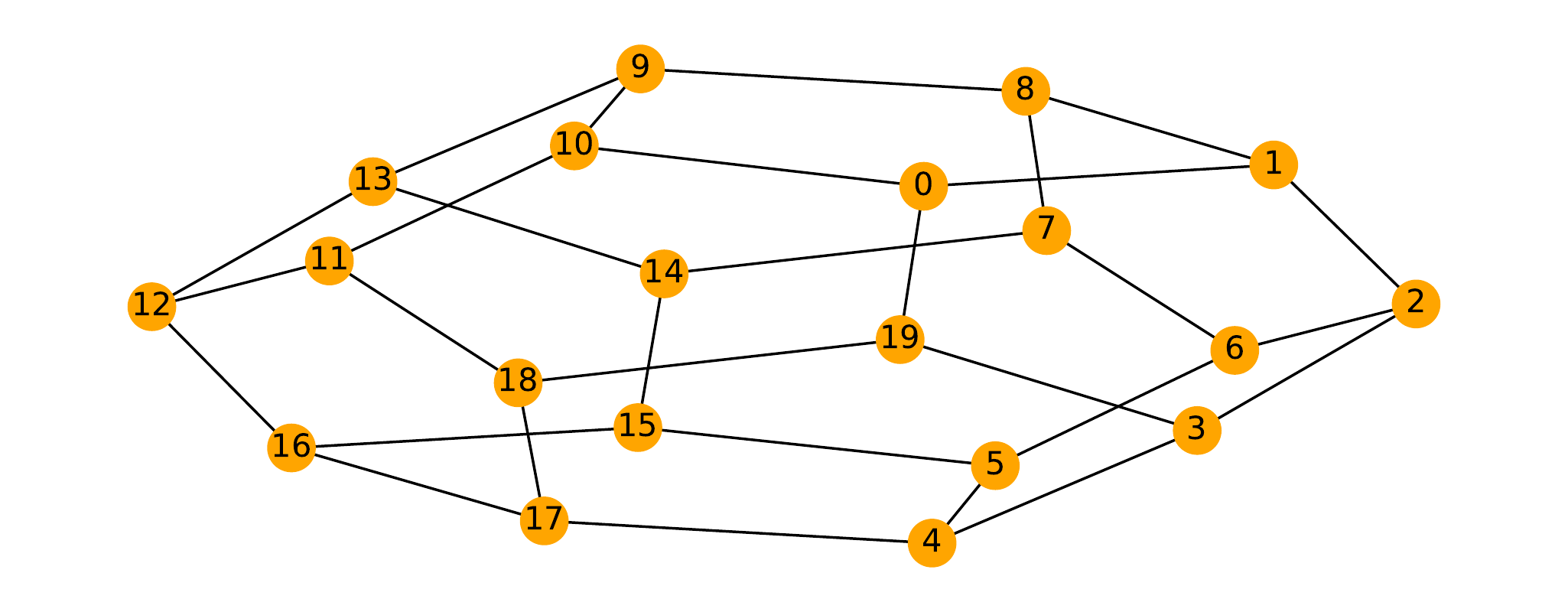}
    \caption{Spherical dependence pattern.}
    \label{fig:spherical_dependence_pattern}
\end{figure}

\begin{figure}[ht]
    \centering
    \includegraphics[width=.9\linewidth]{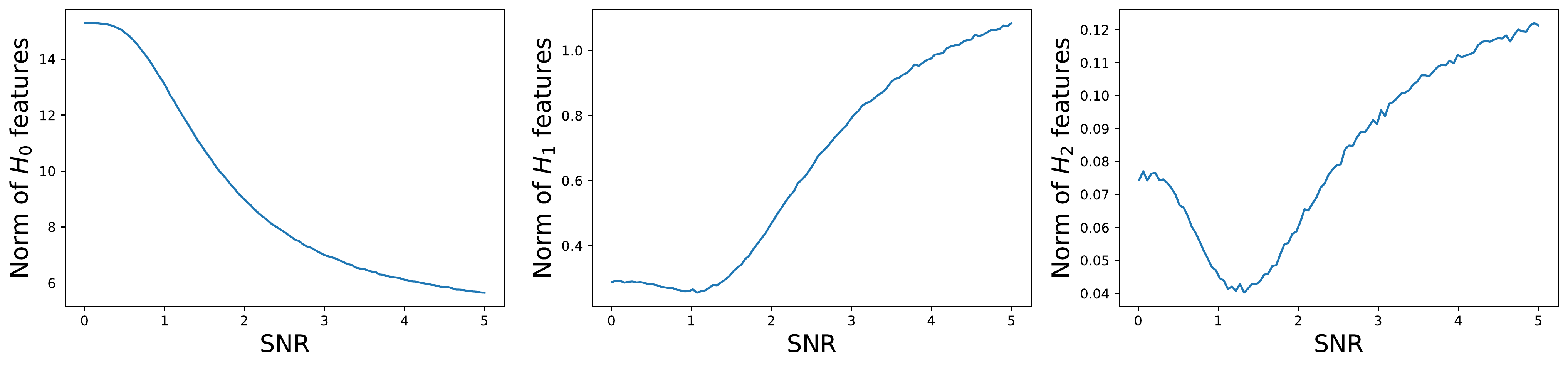}
    \caption{Total persistence as a function of the signal to noise ratio. LEFT: 0-dimensional homology; MIDDLE: 1-dimensional homology; RIGHT: 2-dimensional homology. Total persistence in the $y$-axis and signal to noise ratio $SNR$ in the $x$-axis. The plots are based on the average total persistence for 1000 replicates.}
    \label{fig:sensitivity_analysis}
\end{figure}

In this section we want to study the sensitivity to noise of our approach. Let $Y(t) = S(t) + N(t)$ where $Y(t)$ is the observed signal; $S(t)$ is the underlying signal or stochastic process that is not directly observed whose variance is $\sigma_S^{2}$; $N(t)$ is the additive noise, that is independent of $S(t)$ whose variance is $\sigma_N^{2}$. Then define the signal-to-noise-ratio in terms of the variance of $S(t)$ and $N(t)$ to be $SNR = \frac{\sigma_S^2}{\sigma_N^2}$. To assess the effect of the noise on the topological features of the dependence pattern in the underlying signal, we generate multivariate times series data from a structure that has two dimensional feature. i.e., a spherical structure, see Figure \ref{fig:spherical_dependence_pattern}. 
Let's define the total persistence to be the norm of the persistence diagram' features as follows:
\begin{align}
    P_k = \sum_{i \in PD_k} (d^{k}_i - b^{k}_i)
\end{align}
where $b^{k}_i$ and $d^{k}_i$ represent respectively the birth and death of the $i$-th $k$-dimensional topological feature in the persistence diagram. For every dimension $k$, the total persistence $P_k$ is defined to be the sum of the persistence of all k-dimensional features in the persistence diagram. In what follows, we study the behaviour of the total persistence $P_k$ as a function of the signal to noise ratio, see Figure \ref{fig:sensitivity_analysis}.
The persistence $P_0$ of the 0-dimensional features decreases as the signal to noise ratio grows, which is to be expected because at low SNR, the time series components are mostly independent, resulting in large mutual distances and many unconnected components and at high SNR, the time series components are mostly dependent, resulting in smaller mutual distances, i.e., fewer connected components. On the other hand, the persistence $P_k$ of the 1- and 2-dimensional features increases as the signal to noise ratio grows, which makes sense. At low SNR, the time series components are independent, and the connectivity pattern is not visible, but at high SNR, the time series components are mostly dependent according to the spherical structure, i.e., more 1- and 2-dimensional features.

\section{Application of the proposed simulation methods to statistical inference}

There are many disorders that can alter the connectivity of the brain such as Alzheimer's disease, Parkinson's disease, ADHD etc. It is common for such conditions to alter the topology of the brain's connectivity structure by creating holes, cavities or other patterns in the connectivity network. We propose to illustrate via simulations of multivariate time series how it is possible to discriminate between two topological patterns that differ in their one dimensional homology structure. Based on the idea developed in Sections \ref{subsec:one_main_cycle} and \ref{subsec:two_main_cycle}, we generate $N=50$ samples from one model $\mathcal{M}_1$ with one main cycle in its dependence pattern, and $N=50$ samples from another model $\mathcal{M}_2$ with two main cycles in its dependence pattern:
\begin{align}
    Y^{(1, i)}(t) = W^{(1)} Z^{1, i}(t) + \epsilon^{1, i}(t), i = 1, \hdots, N \\ 
    Y^{(2, i)}(t) = W^{(2)} Z^{1, i}(t) + \epsilon^{2, i}(t), i = 1, \hdots, N
\end{align}
where $W^{(1)}$ and $W^{(2)}$ are respectively the mixing weights for model one and two as defined in Sections \ref{subsec:one_main_cycle} and \ref{subsec:two_main_cycle}, $Z^{1, i}(t)$ and $Z^{2, i}(t)$ are the iid latent processes, $\epsilon^{1, i}(t)$ and $\epsilon^{2, i}(t)$ are the additive Gaussian noise.

After generating the time series for both groups, we compute the corresponding persistence diagrams then we compute a topological summary, total persistence as described in the previous section, i.e., $T^{1}_{i}$ and $T^{2}_{i}$ for $i=1, \hdots, N$. In order to compare the topologies of both groups we compute the group mean of these summaries for the one/two dimensional homology etc., and then assess the variability using a bootstrap approach:
\begin{enumerate}
    \item Draw $T^{1*}_{1}, \hdots, T^{1*}_{N}$ and $T^{2*}_{1}, \hdots, T^{2*}_{N}$ from the empirical distribution based on $T^{1}_{1}, \hdots, T^{1}_{N}$ and $T^{2}_{1}, \hdots, T^{2}_{N}$
    \item Compute the group mean $\widehat{T}^{(1*)}_b = \frac{1}{N}\sum_{i=1}^N T^{1*}_{i}$ and $\widehat{T}^{(2*)}_b = \frac{1}{N}\sum_{i=1}^N T^{2*}_{i}$ for the one and two homology groups.
    \item Repeat B times the previous two steps.
    \item Visualize the boxplot of the bootstrap samples.
\end{enumerate}
\begin{figure}[ht]
    \centering
    \includegraphics[width=.7\linewidth]{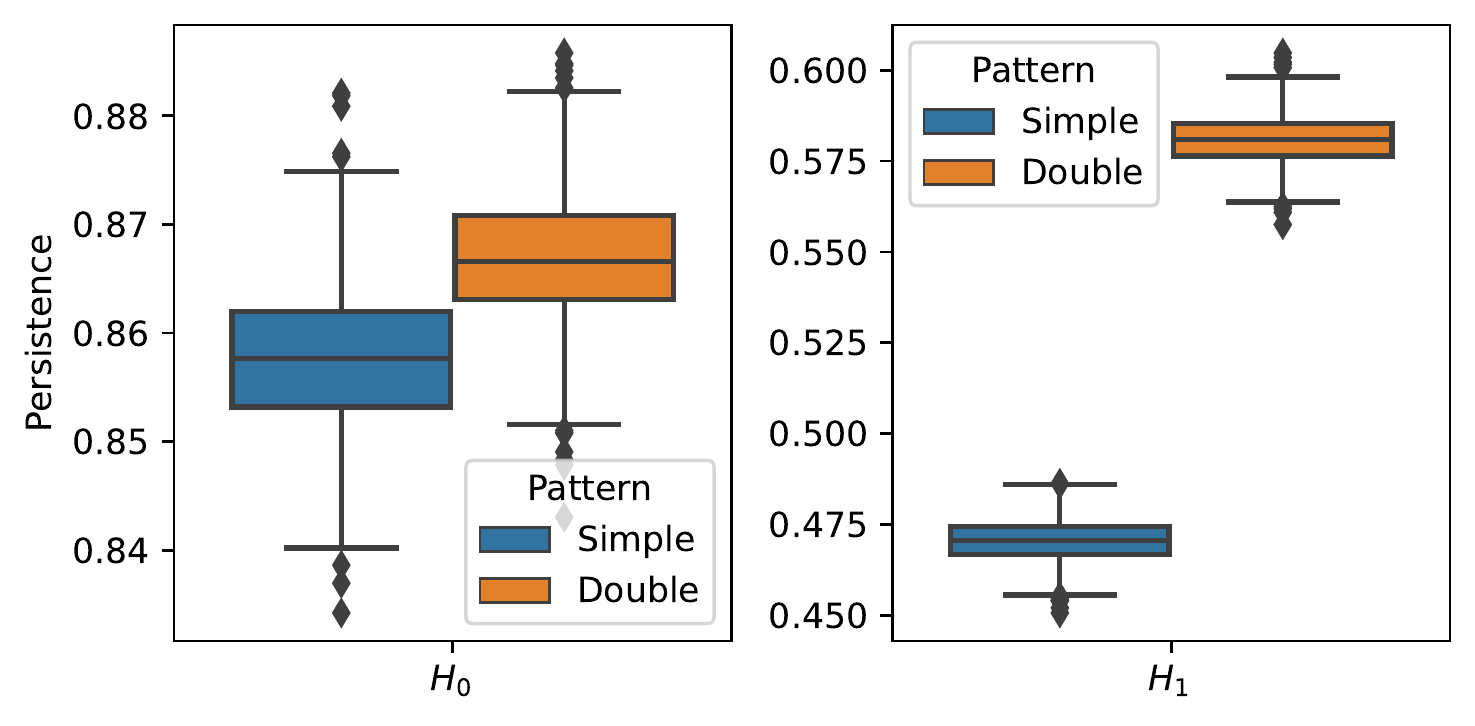}
    \caption{Boxplot of the one (LEFT) and two (RIGHT) homology groups topological summaries for both group simple main cycle (BLUE) and double main cycle (ORANGE) based on $B=1000$ bootstrap samples.}
    \label{fig:simulation_based_inference_boxplot}
\end{figure}
The results of the procedure above can be seen in Figure \ref{fig:simulation_based_inference_boxplot}. It can be seen that the two groups differ mainly in their cyclic structure (1-dimensional homology), high orange boxplot means more persistence of such features but not in their connected components structure (0-dimensional homology). In conclusion, based on the simulated data sets generated from models $\mathcal{M}_1$ and $\mathcal{M}_2$, it is possible to generate multivariate time series data with varying cyclic behaviour in its dependence patterns.

\section{Conclusion}
\label{sec:conclusion}

In this article, we introduce a novel method for simulating multivariate time series data with a predetermined number of cycles in the dependency structure. One primary contribution of the proposed method is to simulate multivariate time series which is essential for assessing the effectiveness of proposed TDA methods. Another major contribution is that our proposed method has advanced TDA on the aspect of statistical inference. To be best of our knowledge, our proposed method is the first that uses mixtures of AR(2) processes in such a way to make the topology of the dependency structure frequency-specific and mimic brain signals. Since our method is fairly general, it may be applied in a wide variety of situations. It can also be utilized to produce higher dimensional topological features. The proposed ideas were illustrated on examples with different cycle counts. A novel procedure based on the quotient group representation to create even more complex dependency patterns such as a torus is presented. To investigate the effect of the variance of the additive noise on the topological features, we conducted. A thorough sensitivity analysis was conducted to study the robustness of our approach. Finally, we gave a demonstration of how our method can be applied to make simulation-based inference.


\bibliographystyle{imsart-nameyear} 
\bibliography{references}   

\end{document}